\documentclass[aps,prd,superscriptaddress,showkeys,showpacs,twocolumn,a4paper]{revtex4-2}
\usepackage{ulem,amssymb}
\usepackage{cancel}
\usepackage{color}
\usepackage{graphicx}
\usepackage{epsfig}
\usepackage{mathrsfs}
\usepackage{amsmath,amsthm}
\usepackage[dvipsnames]{xcolor}
\usepackage{float}
\usepackage[bottom]{footmisc}
\definecolor{darkblue}{RGB}{0,0,196}
\definecolor{darkgreen}{RGB}{0,120,0}
\usepackage[colorlinks=true,linktocpage=true,linkcolor=darkblue,citecolor=red,urlcolor=darkblue]{hyperref}
\usepackage{xcolor}

\newcommand{\be}{\begin{equation}}
\newcommand{\ee}{\end{equation}}
\newcommand{\ba}{\begin{eqnarray}}
\newcommand{\ea}{\end{eqnarray}}

\begin{document}
\title{Electric, thermal and thermoelectric response of a hot pion gas in a time dependent background magnetic field}
\author{Ankit Kumar}
\email{kumar.ankit@iitgn.ac.in}
\affiliation{Indian Institute of Technology Gandhinagar, Gandhinagar-382355, Gujarat, India}
\author{Gowthama K K}
\email{k$_$gowthama@iitb.ac.in}
\affiliation{Indian Institute of Technology Bombay, Mumbai 400076, India}
\author{Vinod Chandra}
\email{vchandra@iitgn.ac.in}
\affiliation{Indian Institute of Technology Gandhinagar, Gandhinagar-382355, Gujarat, India}
\author{Sadhana Dash}
\email{sadhana@phy.iitb.ac.in}
\affiliation{Indian Institute of Technology Bombay, Mumbai 400076, India}
\begin{abstract}
The prime focus of the work is to determine the electric, thermal and thermoelectric transport coefficients of a hot pion gas in the presence of time-dependent background magnetic fields. The thermoelectric effect is analyzed by examining the magneto-Seebeck and Nernst coefficients in the hot pionic medium under such conditions. Furthermore, the phenomenologically relevant elliptic flow coefficient, linked to the Knudsen number, is examined. The analysis reveals the significant impact of both the strength and time dependence of the magnetic field on the transport coefficients of the pionic medium. The results are analyzed in contrast to those obtained under a constant magnetic field.
\end{abstract}

\maketitle
 \section{Introduction}
There is compelling evidence for the formation of a strongly interacting state of matter, known as the Quark Gluon Plasma (QGP) in Heavy-Ion Collision (HIC) experiments at the Relativistic Heavy Ion Collider (RHIC) and Large Hadron Collider (LHC)~\cite{Adams:2005dq,Adcox:2004mh,Back:2004je,Arsene:2004fa,Aamodt:2010pb}. The QGP produced in the collision is expected to achieve thermal and chemical equilibrium before hadronizing into light hadrons. One such important hadron is the $\pi$ meson or the pion.  Pions not only govern the bulk thermodynamic properties of the hadronic phase but also contribute substantially to electromagnetic probes such as soft dileptons and thermal photons\cite{PhysRevC.69.014903,STEELE1996255}. Their abundance and mobility make them key carriers of electric and thermal currents, especially under the influence of external fields. To study the hydrodynamical evolution and properties of a pionic medium, one requires the transport coefficients of the medium as input parameters~\cite{Dash:2020vxk,Das:2019pqd,Kadam:2017iaz}. Understanding the transport properties of pion gas, therefore, becomes essential in bridging theoretical models with experimental data in the post-hadronization regime.

Investigating the influence of electromagnetic fields on the properties of strongly interacting matter has become a central theme in the study of HICs. Non-central collisions at RHIC and LHC are known to generate ultra-strong, transient magnetic fields—of the order of $\sim10^{18-19}$ gauss\cite{KHARZEEV2008227,Skokov:2009qp,doi:10.1142/S0217751X09047570,PhysRevC.107.034901,inghirami2020magnetic}—which strongly influence the evolution and observable signatures of the produced medium. 
A variety of nontrivial phenomena arise in strongly interacting QGP medium in the presence of magnetic fields with strengths comparable to the QCD scale. These include the chiral magnetic effect\cite{PhysRevD.78.074033,Kharzeev:2015znc,Sadofyev:2010pr}, chiral magnetic wave\cite{PhysRevD.83.085007}, modified photon and dilepton production\cite{PhysRevC.88.024910}, magnetic catalysis\cite{Shovkovy2013,Gusynin:1995nb}, and inverse magnetic catalysis\cite{endrHodi2019magnetic}, among others. Such effects significantly modify the properties of the QGP medium and have consequently motivated extensive theoretical and phenomenological investigations.
Magnetic fields of such magnitude, as generated in HICs, are rare in the universe and exist in the interior of certain compact stars known as magnetars\cite{bocquet1995rotatingneutronstarmodels}. Cosmological models also predict that such fields might have existed in the early universe during the electroweak phase transition\cite{PhysRevD.53.662}.

Due to the electrical conductivity of the created med-ium, the magnetic fields produced in the HICs can persist far beyond their initial generation, potentially lasting throughout the QGP phase\cite{PhysRevC.82.034904,PhysRevC.107.034901,Kiril2013ahep,STEWART2021122308}. As the pions are formed in the later stages of the collision, these persistent magnetic fields in the presence of a medium may have a significant impact on the properties of the pion gas. While considerable attention has been devoted to the early QGP phase, the hadronic stage that follows remains equally crucial for interpreting final-state observables. Among hadronic species, pions, being the lightest mesons and the dominant constituents of the hadron gas, play a particularly significant role in this late-time dynamics. This sets motivation for the work presented in the manuscript.    

This study investigates the electrical, thermal, and thermoelectric responses of a hot pion gas in the presence of a time-dependent magnetic field, a scenario more representative of conditions at the RHIC. While previous works have primarily focused on constant background magnetic fields~\cite{Das:2019pqd,Ghosh:2022vjp,PhysRevD.102.076007,Dash:2020vxk,PhysRevD.102.014030,Das:2019wjg}, recent works have looked at the effects of time-dependent magnetic fields on a QGP medium~\cite{Dey:2025hgw,K:2021sct}. We aim to explore how the time-dependent strength of the magnetic field influences charge diffusion, thermal diffusion, and the thermoelectric behavior of the pionic medium. Another interesting exploration is to relate the thermal transport to the elliptic flow parameter through thermal conductivity. The results on the said transport coefficients are seen to be sensitive to the time dependence of the magnetic field in a significant way.

The article is organized as follows. In section~\ref{II}, we derive the electrical conductivity of the pion gas medium in the presence of a time-dependent electromagnetic field to study its electrical response. Section~\ref{III} addresses the thermal response of the medium through thermal transport coefficients and explores the thermoelectric behavior via the magneto-Seebeck and normalized Nernst coefficients. In section~\ref{IV}, we examine the behavior of the Knudsen number and the elliptic flow coefficient. Section~\ref{V} presents a discussion of the results, and finally, section~\ref{VI} concludes the article with a summary of the key findings.

\section{Electrical conductivity of hot pion gas in time-dependent Electromagnetic fields }\label{II}
Pions are found in three forms, $\pi^{\pm}$, $\pi^0$. Among these, only $\pi^{\pm}$ are influenced by the electromagnetic fields generated in the collision. The effect of these fields on the pion gas can be characterized by the electrical conductivity of the medium. The fields cause the distribution function of the pion gas to shift slightly away from the equilibrium, and hence, this shift can be written in terms of the electromagnetic fields and their derivatives, encoding all the information pertaining to the conductivity of the pion gas. To obtain the electrical conductivity, we need to first obtain the expression for the electrical current induced in the medium, which, for a system with non-zero chemical potential $\mu$, is given as
\begin{align}\label{1}
{\bf j}&= \int dP_k\,\epsilon_k\textbf{v}_k\,q_k\,\delta f_k\,, \quad\quad(k=\pi^{\pm}),  
\end{align}
where $\textbf{v}_k=\frac{\textbf{p}_k}{\epsilon_k}$ is the velocity of the particle with $\epsilon_k=\sqrt{p^2_k+m_k^2}$, $dP_k=\frac{d^3\textbf{p}_k}{(2\pi)^3\epsilon_k}$ and $\delta f_k$ is the shift in the distribution function due to external time-dependent electromagnetic fields. $\delta f_k$ can be obtained by solving the Boltzmann equation under the relaxation time approximation. In the presence of electromagnetic fields, the Boltzmann equation takes the following form:
\begin{align}\label{2}
\frac{\partial f_k}{\partial t} +{\bf v}.\frac{\partial f_k}{\partial {\bf x}} +q_{k} [{\bf E} +{\bf v} \times {\bf B}].\frac{\partial f_k}{\partial {\bf p}} =-\frac{\delta f_k}{\tau_R},
\end{align}
where $\tau_R$ is the relaxation time. The relaxation time is evaluated with three species of pions ($\pi^{(\pm)}, \pi^{0}$) for a $2 \rightarrow 2$ process. The momentum averaged relaxation time is dependent on the temperature of the system and is parametrized as, $\tau_R=\Sigma_{i=0}^{3}a_i(\frac{m}{T})^i\frac{1}{T}$, where $a_0=0.0145\,\text{fm}\,\text{GeV}^3$, $a_1=-0.0109\,\text{fm}\,\text{GeV}^3$, $a_2=0.0058\,\text{fm}\,\text{GeV}^3$, and $a_3=0.0026\,\text{fm}\,\text{GeV}^3$, Ref.~\cite{PhysRevD.102.076007}. We adapt the following ansatz for the $\delta f_k$ to solve the Boltzmann equation,
\begin{equation}\label{3}
\delta f_k=({\textbf{p}_k}.{\bf \Xi} ) \frac{\partial f^0_k}{\partial \epsilon_k}, 
\end{equation} 
where ${\bf{\Xi}}$ carries the information of the electromagnetic fields and their leading order space and time derivatives, in the form as follows,
\begin{align}\label{4}
\mathbf{\Xi} =& \alpha_1\textbf{E}+ \alpha_2\dot{\textbf{E}}+ \alpha_3(\textbf{E}\times \textbf{B})+ \alpha_4(\dot{\textbf{E}}\times \textbf{B})+ \alpha_5(\textbf{E}\times \dot{\textbf{B}})\nonumber\\&+\alpha_6 ({\pmb \nabla} \times \textbf{E}) +\alpha_7 \textbf{B}+\alpha_8 \dot{\textbf{B}}+\alpha_9 ({\pmb \nabla} \times \textbf{B}).
\end{align} 
Here, $\alpha_{i}$ ($i=1, 2,.., 9$) denote functions that are to be determined and correspond to the electric charge transport coefficients. These functions can be derived from a microscopic description of the pion gas. In this study, we consider the case where the chiral chemical potential vanishes, which implies $\alpha_{i}=0$ for $i=(6,7,8)$ since parity is conserved~\cite{Satow:2014lia}. 
Using Eqs.~(\ref{3}) and (\ref{4}) in Eq.~(\ref{2}) we get,
\begin{widetext}
\begin{align}\label{6}
& \epsilon_k {\bf v}_k.\Big[\alpha_1 \dot{{\bf E}}+\dot{\alpha_1 }{\bf E}+\alpha_2 \ddot{{\bf E}}+\dot{\alpha_2 }\dot{{\bf E}}+\alpha_3 (\dot{{\bf E}}\times {\bf B}) +\alpha_3 ({\bf E}\times \dot{{\bf B}})+\dot{\alpha_3 }({\bf E}\times {\bf B})+\alpha_4 (\dot{{\bf E}}\times \dot{{\bf B}})+\alpha_4 (\ddot{{\bf E}} \times {\bf B}) +\dot{\alpha_4 }(\dot{{\bf E}} \times {\bf B}) \nonumber\\
&+\alpha_5 (\dot{{\bf E}}\times \dot{{\bf B}})+\alpha_5 ({\bf E}\times \ddot{{\bf B}})+\dot{\alpha_5 }({\bf E}\times \dot{{\bf B}})+\alpha_9 (\pmb{\nabla} \times {\bf \dot{B}})+\dot{\alpha_9}(\pmb{\nabla} \times {\bf B})\Big]
 +q_{k}{\bf v}_k.{\bf E}-\alpha_1 q_{k}{{\bf v}_k}.({\bf E}\times {\bf B})
 -\alpha_2 q_{k}{\bf v}_k.(\dot{{\bf E}}\times {\bf B})\nonumber\\
 &+\alpha_3 q_{k}({\bf v}_k.{\bf E})(B^2)-\alpha_3 q_{k}({\bf v}_k.{\bf B})({\bf B}.{\bf E})+\alpha_4 q_{k}({\bf v}_k.\dot{{\bf E}})(B^2)-\alpha_4 q_{k}({\bf v}_k.{\bf B})({\bf B}.\dot{{\bf E}})+\alpha_5 q_{k}({\bf v}_k.{\bf E})(\dot{{\bf B}}.{\bf B})-\alpha_5 q_{k}(\dot{{\bf B}}.{\bf v}_k)({\bf E}.{\bf B})\nonumber\\
 &-\alpha_9 q_{k}({\bf B.v}_k)(\pmb{\nabla}.{\bf B})=-\frac{\epsilon_k}{\tau_R}\Big[\alpha_1 {\bf v}_k.{\bf E}+\alpha_2 {\bf v}_k.\dot{{\bf E}}+\alpha_3 {\bf v}_k.({\bf E}\times {\bf B})+\alpha_4 {\bf v}_k.(\dot{{\bf E}}\times {\bf B})+\alpha_5 {\bf v}_k.({\bf E}\times \dot{{\bf B}})+\alpha_9 {\bf v}_k.(\pmb{\nabla} \times {\bf B})\Big].
\end{align}
\end{widetext}

Here, we have ignored the terms having derivatives of the fields higher than the first order in space-time, assuming that the fields vary slowly with space and time. This assumption allows for the incorporation of collisional effects in the pion gas. Therefore, the terms with $\dot{\alpha_2}, \dot{\alpha_4}, \dot{\alpha_5}, \dot{\alpha_9}$ vanish. Using this in Eq.~\eqref{6} and comparing the coefficients of the tensorial structures on both sides, we get the following coupled differential equations:
\begin{align}\label{9}
  & \dot{\alpha_1} =  -\bigg[\frac{1}{\tau_R}\alpha_1 +\Big(\frac{q_{k}B^2}{\epsilon_k} -\frac{\tau_R q_{k}B\dot{B}}{\epsilon_k}\Big)\alpha_3 +\frac{q_{k}}{\epsilon_k}\bigg],\nonumber \\  
  & \dot{\alpha_3} = -\frac{1}{\tau_R}\alpha_3 +\frac{q_{k}}{\epsilon_k}\alpha_1,
\end{align}
along with the coupled equations
\begin{align}\label{7}
&\alpha_2=-\tau_R\Big[ \alpha_1 +\frac{q_{k} \alpha_4 B^2}{\epsilon_k}\Big],
 &&\alpha_5 = -\tau_R \alpha_3,\nonumber \\
     &{\alpha_4}=-{\tau_R}\Big[\alpha_3-\frac{\alpha_2 q_{k}}{\epsilon_k}\Big].
\end{align}
Here, $\dot{B}=|{\bf \dot{B}}|$. The coupled differential equations can be written in the form of a matrix equation as
\begin{equation}\label{12}
 \frac{d X}{d t}= AX +G,   
\end{equation}
where the matrices $X$, $A$ and $G$ are
\begin{align}
 X=\begin{pmatrix}
\alpha_1\\\alpha_3
\end{pmatrix},  
&& A=\begin{pmatrix}
 -\frac{1}{\tau_R} & -\frac{q_{k} F^2}{\epsilon_k}\\
\frac{q_{k}}{\epsilon_k} &  -\frac{1}{\tau_R}, 
\end{pmatrix},
&&& G=\begin{pmatrix}
\frac{-q_{k}}{\epsilon_k}\\ 0
\end{pmatrix},\nonumber
\end{align}
with $F = \sqrt{B(B-\tau_R \dot{B})}$. To solve Eq.~(\ref{12}), we employ the method of variation of constants. We first diagonalize the matrix $A$ and solve the homogeneous equation $\frac{dX}{dt}=A X$. Then, we consider Eq.~(\ref{12}) with its non-homogeneous part and solve it by promoting the constants of integration to time-dependent functions $k_1(t)$ and $k_2(t)$. The resulting particular solution is
\begin{align}\label{16}
&\alpha_1 =k_1 iFe^{\eta_1} -k_2 iFe^{\eta_2},
&&\alpha_3=k_1 e^{\eta_1} +k_2 e^{\eta_2}. 
\end{align}
The functions $k_1(t)$ and $k_2(t)$ take the form $k_1 =\frac{iq_{k}}{2\epsilon_k} I_1$ and  $k_2 =-\frac{iq_{k}}{2\epsilon_k} I_2$ where $\eta_j$ and $I_j$, $(j=1, 2)$ are
\begin{align}\label{21}
    &\eta_j = -\frac{t}{\tau_R} +a_j\frac{q_{k} i}{\epsilon_k}\int F dt,
    &&I_j = \int \frac{ e^{-\eta_j}}{F},
\end{align}
with $a_1=1$ and $a_2=-1$. Using Eq.~\eqref{21} in Eq.~(\ref{16}) and using Eq.~(\ref{7}), we get the expressions for $\alpha_i$ as
\begin{align}\label{20}
 &\alpha_1 =-\frac{\Tilde{\Omega}_k}{2}(I_1 e^{\eta_1} +I_2 e^{\eta_2} ),\\
 &\alpha_3 =\frac{q_{k}i}{2\epsilon_k}(I_1 e^{\eta_1} -I_2 e^{\eta_2} ),\label{20.2}\\
 &\alpha_5 =-\frac{\tau_R q_{k}i}{2\epsilon_k}(I_1 e^{\eta_1} -I_2 e^{\eta_2}  ),\label{20.4}\\
 &\alpha_2 =\frac{(\frac{\Tilde{\Omega}_k\tau_R}{2} +\frac{i\Omega^2_k\tau^2_R}{2})I_1 e^{\eta_1}+(\frac{\Tilde{\Omega}_k\tau_R}{2}-\frac{i\Omega^2_k\tau^2_R}{2}) I_2 e^{\eta_2}}{1+\Omega^2_k\tau^2_R}\label{20.1}\\
 &\alpha_4 =\frac{(\frac{\Tilde{\Omega}^2_k\tau^2_R}{2F} -\frac{iq_{k}\tau_R}{2\epsilon_k})I_1 e^{\eta_1}+(\frac{\Tilde{\Omega}^2_k\tau^2_R}{2F}+\frac{iq_{k}\tau_R}{2\epsilon_k}) I_2 e^{\eta_2}}{1+\Omega^2_k\tau^2_R},\label{20.3}
\end{align}
where $\Omega_k=\frac{q_{k}B}{\epsilon_k}$ is the cyclotron frequency and $\Tilde{\Omega}_k=\frac{q_{k}F}{\epsilon_k}$. Now, we solve the $\alpha_i$ for the time-dependent electromagnetic fields. We consider the electric and magnetic field magnitudes of the form $E=E_0 e^{-t/\tau_E}$ and $B=B_0 e^{-t/\tau_B}$, where $\tau_E$ and $\tau_B$ are the decay parameters of electric and magnetic fields respectively. Theoretical modeling of the decay of the magnetic field in a conducting medium like QGP indicates that the fields may persist beyond the hadronization of the medium\cite{PhysRevC.107.034901,Kiril2013ahep}. As the pion gas is formed in the later stages, after hadronization of the medium, we consider a very slowly decaying magnetic field and take the large values of $\tau_B$ and $\tau_E$ in our analysis.

The induced current can then be decomposed into Ohmic and Hall currents as $\textbf{j}=j_e\hat{\textbf{e}}+j_H(\hat{\textbf{e}}\times\hat{\textbf{b}})$, with
\begin{align}
    j_e=j_e^{(0)}+j_e^{(1)}, \quad\quad\quad j_H=j_H^{(0)}+j_H^{(1)}+j_H^{(2)},
\end{align}
where $j_e^{(0)}$ is the Ohmic current and $j_e^{(1)}$ is the correction to the Ohmic current due to the time dependence of the electromagnetic fields. Similarly, $j_H^{(0)}$ is the Hall current and $j_H^{(1)}$ and $j_H^{(2)}$ are the corrections to it because of the $(\dot{\textbf{E}}\times\textbf{B})$ and $(\textbf{E}\times\dot{\textbf{B}})$ terms in Eq.~\eqref{4}, respectively. These currents are given as
\begin{align}
    j_e^{(0)}&=\frac{Ee}{3}\int dP\,p^2\bigg\{\big(\alpha_1\frac{\partial f^0}{\partial\epsilon}\big)_{\pi^+}-\big(\alpha_1\frac{\partial f^0}{\partial\epsilon}\big)_{\pi^-}\bigg\}, \\
    j_e^{(1)}&=\frac{\dot{E}e}{3}\int dP\,p^2\bigg\{\big(\alpha_2\frac{\partial f^0}{\partial\epsilon}\big)_{\pi^+}-\big(\alpha_2\frac{\partial f^0}{\partial\epsilon}\big)_{\pi^-}\bigg\}, \\
    j_H^{(0)}&=\frac{EBe}{3}\int dP\,p^2\bigg\{\big(\alpha_3\frac{\partial f^0}{\partial\epsilon}\big)_{\pi^+}-\big(\alpha_3\frac{\partial f^0}{\partial\epsilon}\big)_{\pi^-}\bigg\}, \\
    j_H^{(1)}&=\frac{\dot{E}Be}{3}\int dP\,p^2\bigg\{\big(\alpha_4\frac{\partial f^0}{\partial\epsilon}\big)_{\pi^+}-\big(\alpha_4\frac{\partial f^0}{\partial\epsilon}\big)_{\pi^-}\bigg\}, \\
    j_H^{(2)}&=\frac{E\dot{B}e}{3}\int dP\,p^2\bigg\{\big(\alpha_5\frac{\partial f^0}{\partial\epsilon}\big)_{\pi^+}-\big(\alpha_5\frac{\partial f^0}{\partial\epsilon}\big)_{\pi^-}\bigg\},
\end{align}
where $e$ is the electric charge on the pion. Since the three species of pions have almost the same mass, we take $\frac{p^2}{\epsilon}$ outside the curly brackets above. The Ohmic and Hall conductivities are given as $\sigma_e=\frac{j_e^{(0)}+j_e^{(1)}}{E}$ and $\sigma_H=\frac{j_H^{(0)}+j_H^{(1)}+j_H^{(2)}}{E}$ respectively.
 
\section{Thermal and thermoelectric response of pion gas}\label{III}

\subsection*{Thermal response}

The energy-momentum tensor $T^{\mu\nu}$ and particle flow $N^{\mu}$ of the hot pion gas can be defined in terms of their momentum distribution function as follows,
\begin{align}\label{1.1}
T^{\mu\nu}(x)=\int{dP_k\,{p}_k^{\mu}\,{p}_k^{\nu}\,f_k(x,{p}_{k})},
\end{align}
and
\begin{align}\label{1.2}
N^{\mu}(x)=\int{dP_k\,{p}_k^{\mu}\,f_k(x,{p}_{k})},
\end{align}
respectively. Employing the form of the distribution function, $f_k=f^0_k+\delta f_k$, in Eq.~(\ref{1.1}) and Eq.~(\ref{1.2}),  the total macroscopic quantities can be split into the equilibrium parts, $T^{0~\mu\nu}, N^{0~\mu}$, and the away from equilibrium parts, $\Delta T^{\mu\nu} (x), \Delta N^{\mu} (x)$. The thermal response of the pion gas arises due to the non-equilibrium part of the system and can be studied in terms of dissipative net heat flow. The heat current utilizing these quantities can be written as, 
\begin{align}\label{1.4}
{ I^i_k}= \Delta T^{0i}_k -h \Delta N^i_k,  
 \end{align}
where $h$ is the enthalpy per particle given by $h=\frac{\varepsilon -P}{n}$, with $P$ being the pressure, $\varepsilon$ the energy density, and $n$ the number density, each defined as in \cite{PhysRevD.97.034032}. 
Employing Eq.~(\ref{1.1}) and Eq.~(\ref{1.2}), the microscopic definition of heat flow takes the form as,
\begin{align}\label{1.5}
   {\bf I}_k = \sum_k \int dP_k \,{\bf p}_k (\epsilon_k -h)\delta f_k.
\end{align}
Solving the relativistic Boltzmann equation with RTA type collision kernel as in Eq. (\ref{2}), the shift in the distribution function, $\delta f_k$, can be written as 
\begin{equation}\label{1.10}
\delta f_k=({\bf{p}}.{\bf \Xi} ) \frac{\partial f^0_k}{\partial \epsilon_k}.
\end{equation} 
As the system is driven away from equilibrium due to the external time dependent magnetic field and the local thermal driving force $\textbf{X}$ defined as $X_{i}=\frac{\partial_i T}{T}-\frac{\partial_i P}{n h}$\cite{groot1980relativistic}, the vector $\mathbf{\Xi}$ can be written as 
\begin{align}\label{5.9}
\mathbf{\Xi} =& \beta_1\textbf{B}+ \beta_2\textbf{X}+ \beta_3(\textbf{X}\times \textbf{B})+\beta_4 \dot{\textbf{B}} +\beta_5(\textbf{X}\times \dot{\textbf{B}}).
\end{align} 
In our analysis, we consider slowly decaying magnetic fields, and hence higher-order derivative terms are neglected. We now follow the same steps as in section \ref{II} to obtain the coefficients $\beta_i$ ($i=1\,\,\text{to}\,\,5$).
The detailed calculation of $\beta_i$ is given in \cite{PhysRevD.106.034008}.  
The  general forms of the coefficients $\beta_i$ are obtained as follows,  
\begin{align}\label{11}
 &\beta_1 =-\frac{i(\textbf{B}.\textbf{X})}{F \epsilon_k} I_0 e^{\eta_0}+\frac{i(\textbf{B}.\textbf{X})}{2F \epsilon_k} I_1 e^{\eta_1} +\frac{i(\textbf{B}.\textbf{X})}{2F \epsilon_k} I_2 e^{\eta_2},\\
 &\beta_2 =(\epsilon_k-h)\frac{F}{2 \epsilon_k} I_1 e^{\eta_1}+(\epsilon_k-h)\frac{F}{2 \epsilon_k} I_2 e^{\eta_1} ,\label{20.2}\\
 &\beta_3 =i(\epsilon_k-h)\frac{ I_1}{2 \epsilon_k} e^{\eta_1} -i(\epsilon_k-h)\frac{I_2}{2 \epsilon_k} e^{\eta_1},\label{20.4}\\
 &\beta_4 = -\tau_R \beta_1 -\frac{q_{k} \tau_R^2}{\epsilon_k} \beta_3 (\textbf{B}.\textbf{X}),\label{20.1}\\
 &\beta_5 =-\tau_R \beta_3\label{11.11},
\end{align}

\begin{figure*}
    \centering
    \centering
    \hspace{-2.5cm}
    \includegraphics[width=0.405\textwidth]{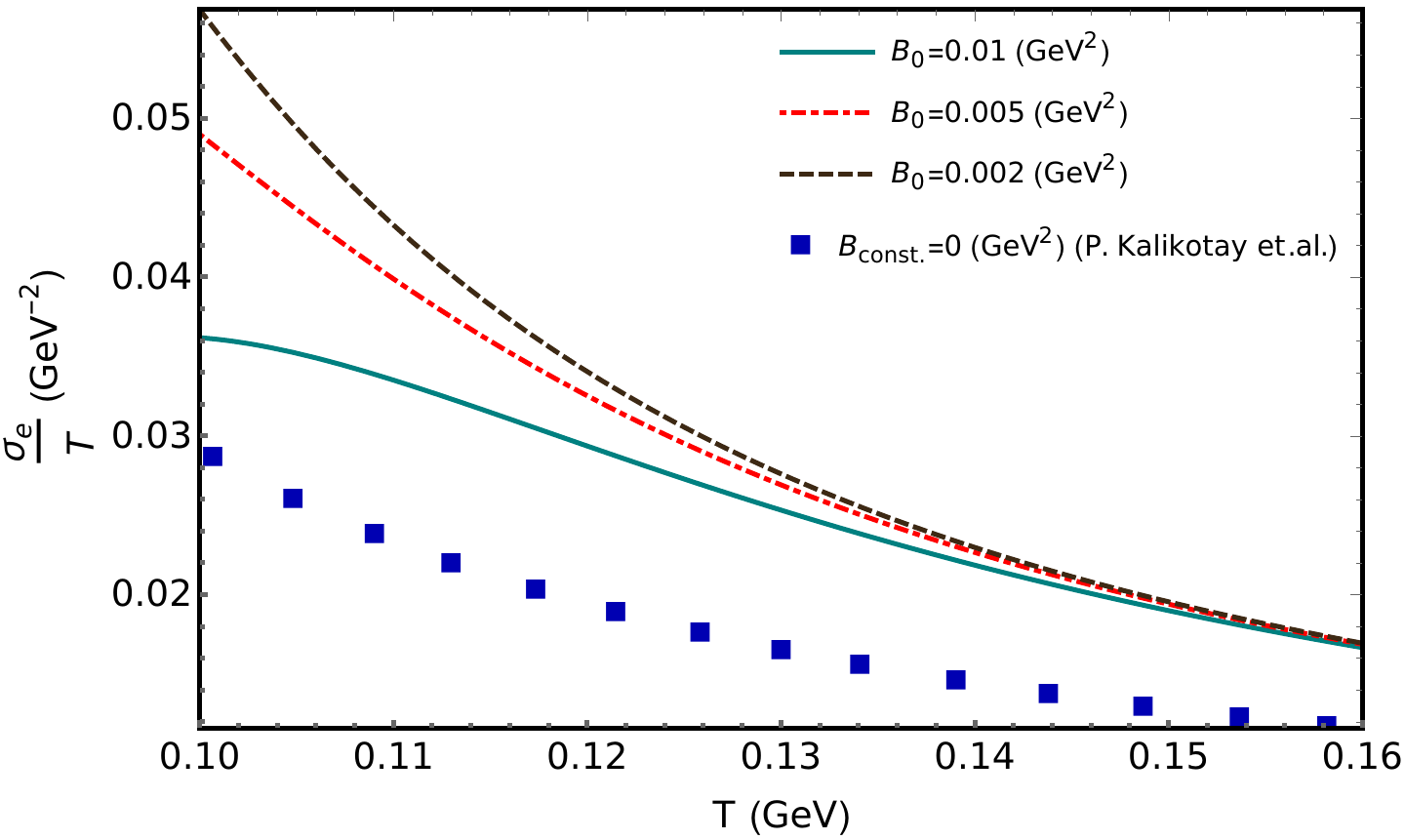}
    \hspace{0.2cm}
    \includegraphics[width=0.405\textwidth]{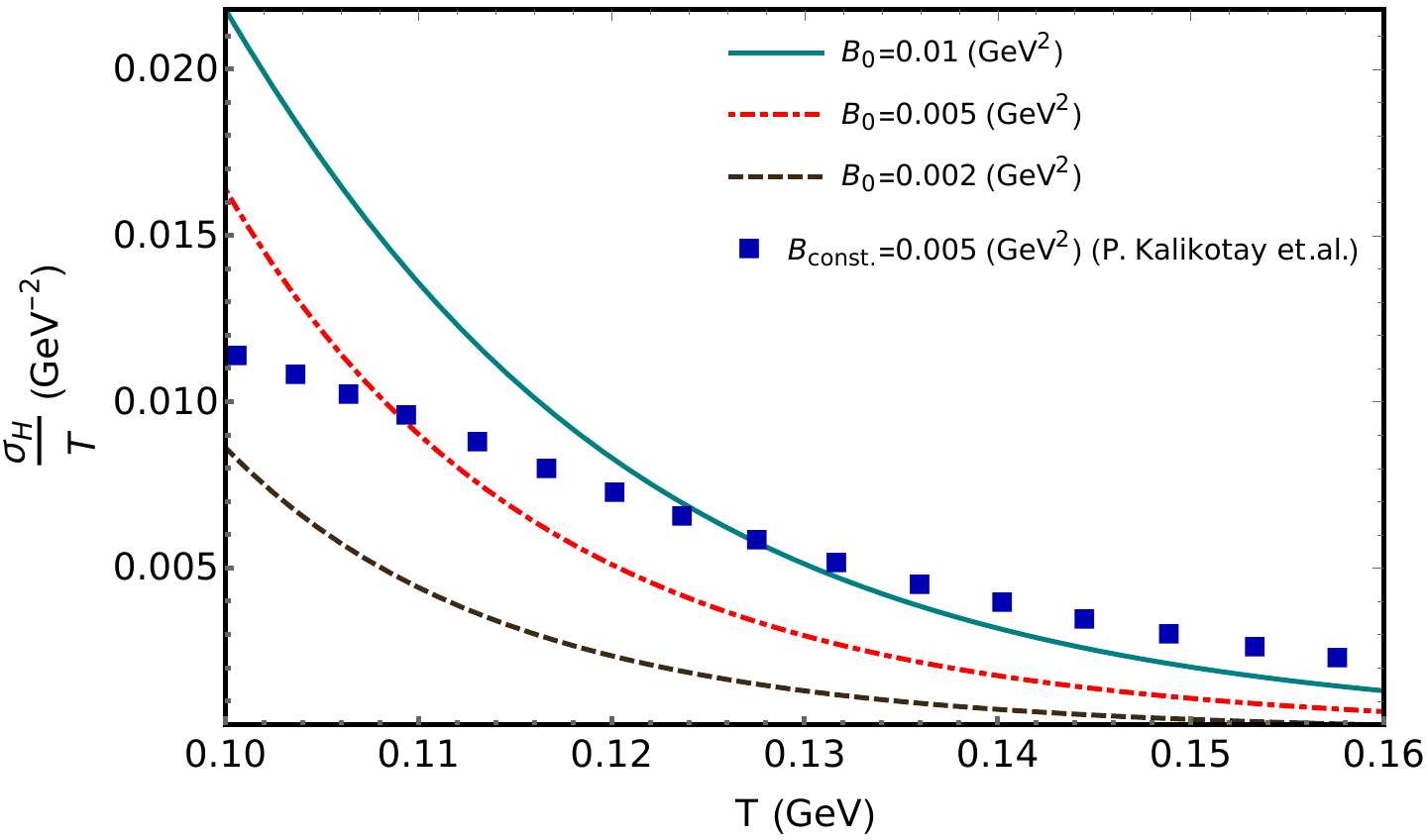}
    \hspace{-2.5cm}
    \caption{\small Ohmic conductivity (left panel) and Hall conductivity (Right panel) as a function of temperature for different amplitudes of magnetic field at $t=2.5\, \text{fm}$ and $(\mu)_{\pi^{\pm}}=\pm0.1\,\text{GeV}$. The results have been compared with \cite{PhysRevD.102.076007}.  }
\label{f1}
\end{figure*}

\begin{figure*}
    \centering
    \centering
    \hspace{-2.5cm}
    \includegraphics[width=0.405\textwidth]{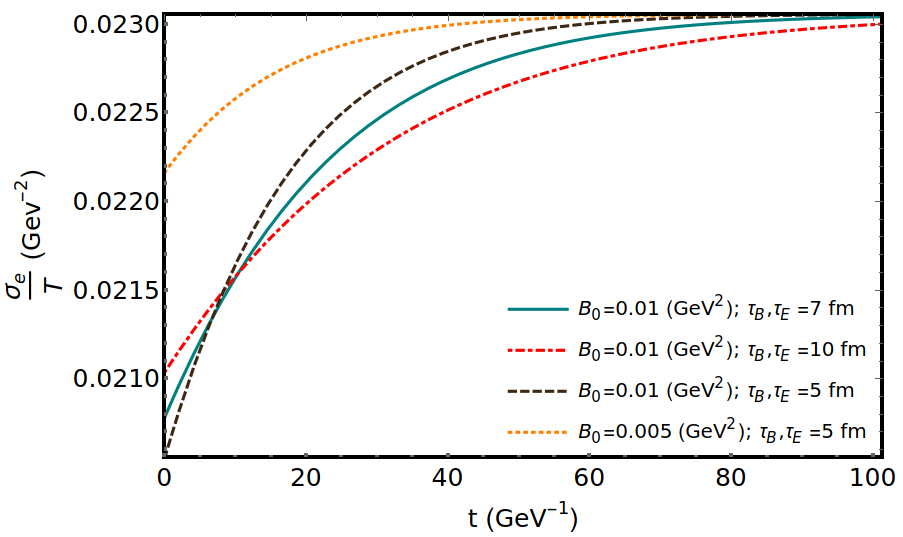}
    \hspace{0.2cm}
    \includegraphics[width=0.405\textwidth]{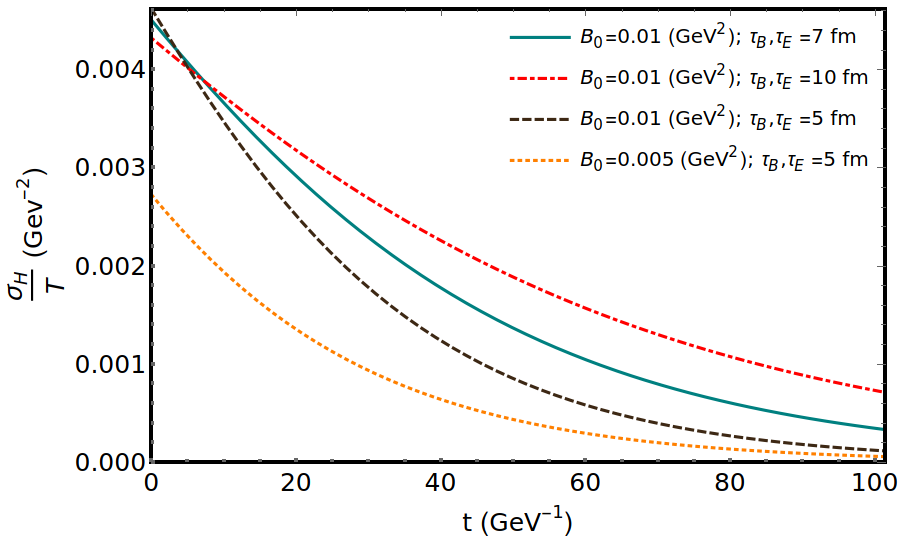}
    \hspace{-2.5cm}
    \caption{\small Ohmic conductivity (left panel) and Hall conductivity (Right panel) as a function of time for different amplitudes and decay parameters of magnetic and electric fields at $T=0.14\,\text{GeV}$ and $(\mu)_{\pi^{\pm}}=\pm0.1\,\text{GeV}$.  }
\label{f2}
\end{figure*}

\begin{figure*}
    \centering
    \centering
    \hspace{-2.5cm}
    \includegraphics[width=0.405\textwidth]{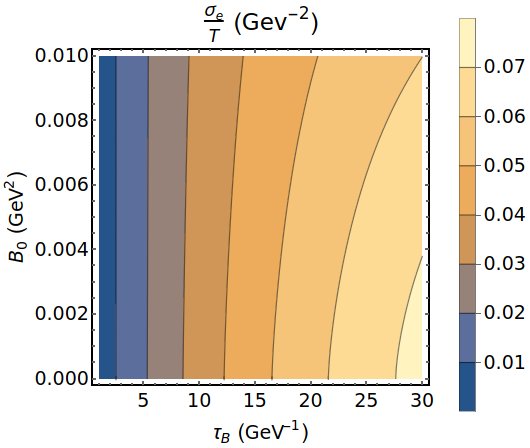}
    \hspace{0.2cm}
    \includegraphics[width=0.405\textwidth]{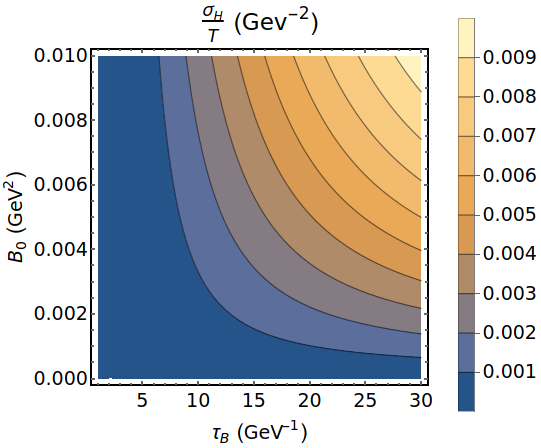}
    \hspace{-2.5cm}
    \caption{\small Ohmic conductivity (left panel) and Hall conductivity (Right panel) as a function of amplitude and decay parameter of the magnetic field at $t=2.5\, \text{fm}$, $T=0.12\,\text{GeV}$ and $(\mu)_{\pi^{\pm}}=\pm0.1\,\text{GeV}$.  }
\label{f3}
\end{figure*}

\begin{figure*}
    \centering
    \centering
    \hspace{-2.5cm}
    \includegraphics[width=0.405\textwidth]{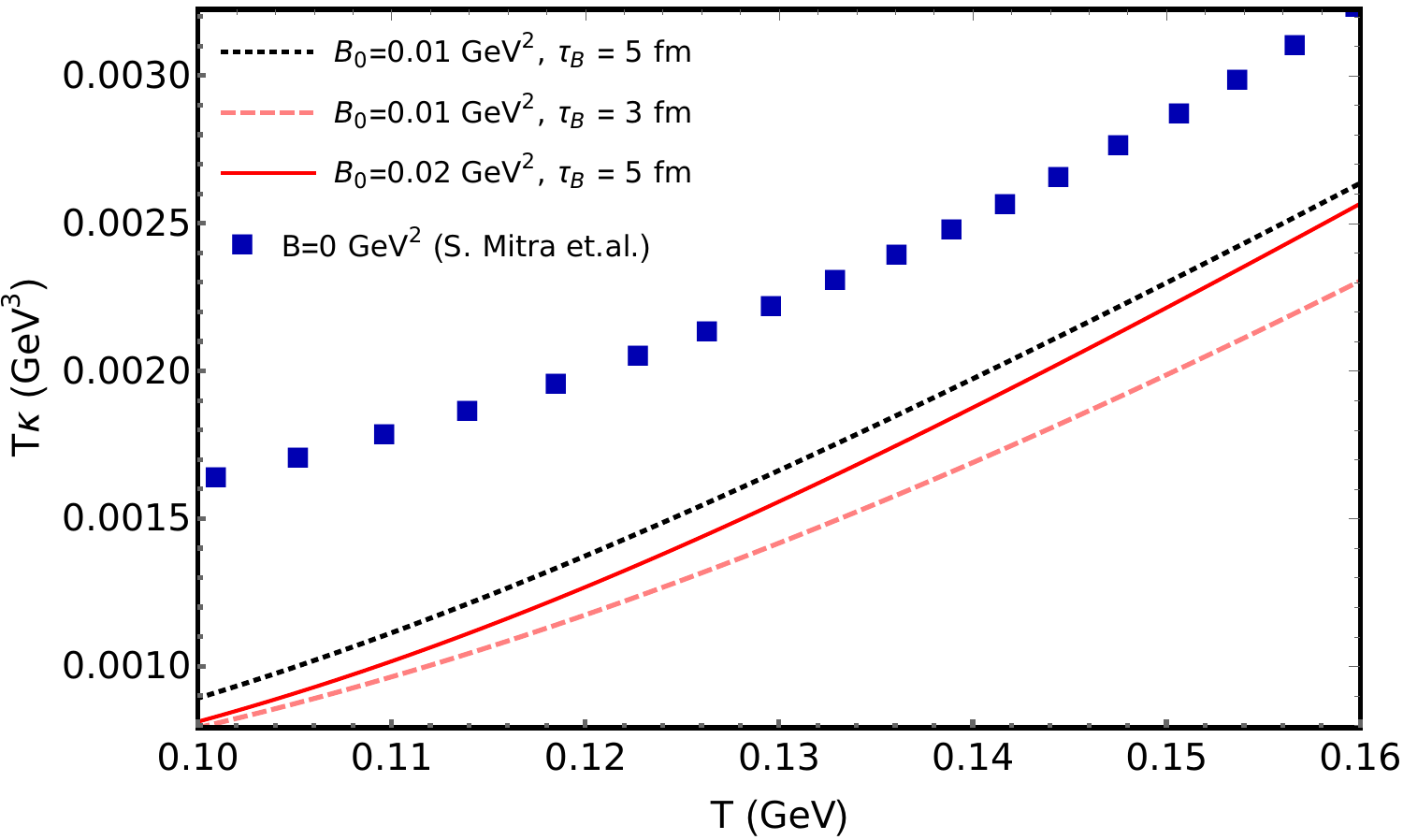}
    \hspace{0.2cm}
    \includegraphics[width=0.405\textwidth]{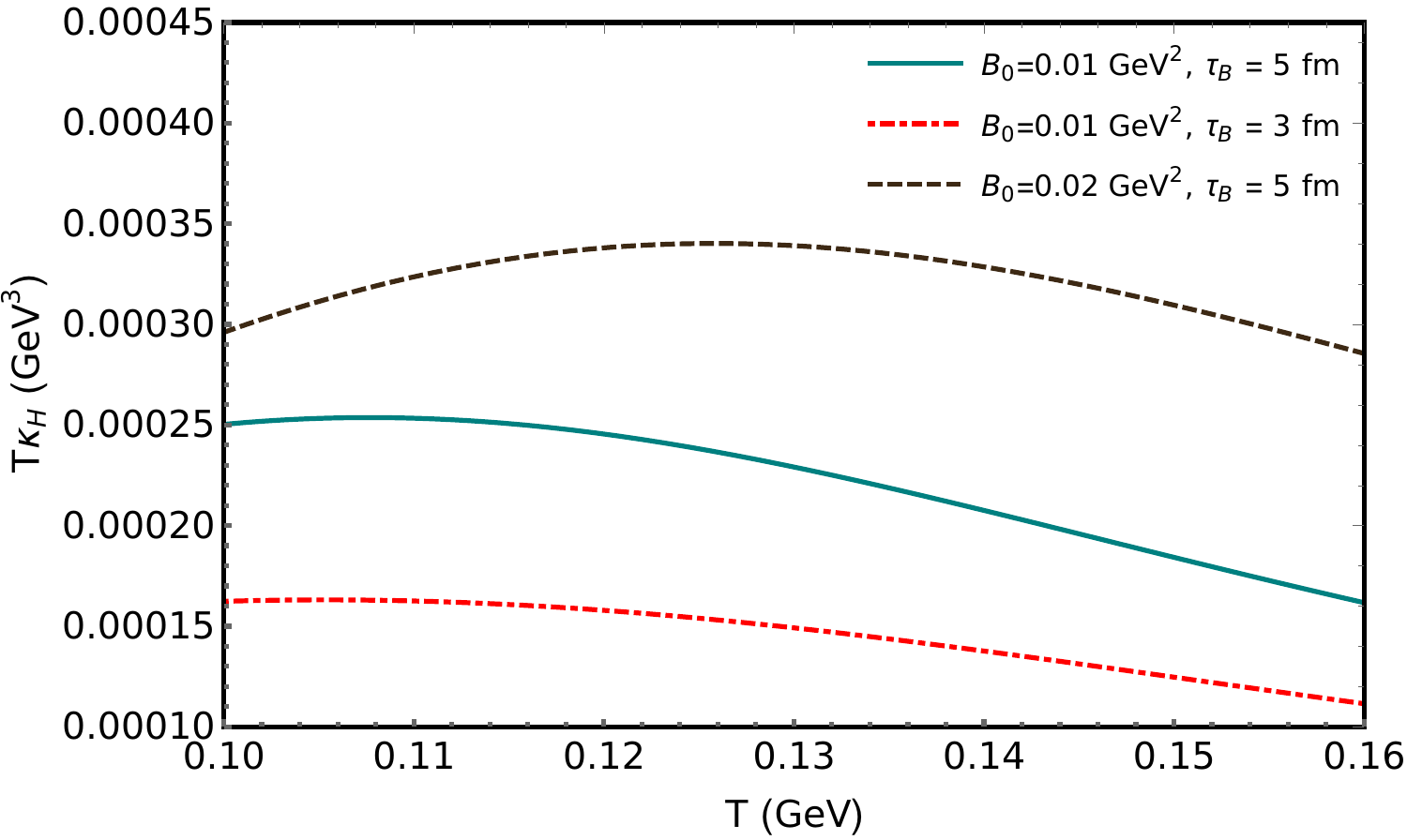}
    \hspace{-2.5cm}
    \caption{\small Thermal conductivity (left panel) and Hall-like Thermal conductivity (right panel) as a function of temperature for different values of amplitude and decay parameter of the magnetic field at $t=2.5\, \text{fm}$, $(\mu)_{\pi^{\pm}}=\pm0.1\,\text{GeV}$ and $(\mu)_{\pi^{0}}=0\,\text{GeV}$. The results of the thermal conductivity have been compared with \cite{PhysRevD.89.054013}. }
\label{f4}
\end{figure*}
where the functions $\eta_j$ and $I_j$ are as defined in Eq.~(\ref{21}) with $a_0=0$. They depend on the profile of the magnetic field evolution through $F$. In the rest of the analysis, the coefficients, $\beta_{1}$ and $\beta_4$ are neglected assuming parity symmetry and choosing $\textbf{B}.\textbf{X}=0$. Further, employing Eq.~(\ref{1.10}) with \eqref{5.9} in Eq.~(\ref{1.5}), the heat current of the pion gas in the presence of time time-decaying magnetic field takes the form: 
\begin{align}\label{1.111}
\textbf{I} = -\kappa T \textbf{X} +\big(\bar{\kappa_1}+ \bar{\kappa_2}\big) T  (\textbf{X} \times \textbf{b}),  
\end{align}
here $\kappa$, $\bar{\kappa_1}$, and $\bar{\kappa_2}$ are the thermal transport coefficients and $\textbf{b}$ is the unit vector of the magnetic field. The first term in Eq.~(\ref{1.111}) denotes the heat flow in the direction of the thermal gradient and is characterized by the thermal conductivity, $\kappa$,
\begin{align}\label{18}
    \kappa &=  -\frac{1}{3T} \int dP\,p^2(\epsilon -h) \bigg\{\big(\beta_2\frac{\partial f^0}{\partial \epsilon}\big)_{\pi^+}\\ \nonumber
    &\quad\quad\quad+\big(\beta_2\frac{\partial f^0}{\partial \epsilon}\big)_{\pi^-}\bigg\}-(\kappa)_{\pi^0}, 
\end{align}
where $(\kappa)_{\pi^0}$ is the contribution of $\pi^0$ to the thermal conductivity of the pion gas, and is not influenced by the magnetic field, and is the same as in literature\cite{PhysRevD.89.054013}. It is given as,
\begin{align}
    (\kappa)_{\pi^0}=\frac{1}{3T} \int dP\,\frac{p^2}{\epsilon}\,(\epsilon -h)^2\,\tau_R\,\big(\frac{\partial f^0}{\partial \epsilon}\big)_{\pi^0}.
\end{align}
The effects of time dependence of the magnetic field enter Eq.~\eqref{18} through $\beta_2$. Eq.~\eqref{18} reduces to the forms in Ref.~\cite{PhysRevD.96.094003} and Ref.~\cite{Das:2019pqd} in the limits of $B=0$ and of constant magnetic field, respectively. The coefficients $\bar{\kappa}_1$ and $\bar{\kappa}_2$ represent Hall-like thermal conductivities that arise purely due to the magnetic field and are of the form, 
\begin{align}\label{}
    &\bar{\kappa}_1 = \frac{B}{3T}  \int dP\,p^2(\epsilon -h) \bigg\{\big(\beta_3\frac{\partial f^0}{\partial \epsilon}\big)_{\pi^+}+\big(\beta_3\frac{\partial f^0}{\partial \epsilon}\big)_{\pi^-}\bigg\},\label{23.1} \\
     &\bar{\kappa}_2 = \frac{\dot{B}}{3T}  \int dP\,p^2(\epsilon -h) \bigg\{\big(\beta_5\frac{\partial f^0}{\partial \epsilon}\big)_{\pi^+}+\big(\beta_5\frac{\partial f^0}{\partial \epsilon}\big)_{\pi^-}\bigg\}.\label{}
\end{align}
The presence of a Lorentz force term in the relativistic Boltzmann equation due to an external magnetic field changes the behavior of the charged particles in the medium. As in the case with charge currents, an additional current perpendicular to the magnetic field and the thermal driving force is generated. This phenomenon is characterized by the coefficient, $\bar{\kappa}_1$, and $\bar{\kappa}_2$ is the correction to $\bar{\kappa}_1$ due to the time dependent nature of the magnetic field. $\bar{\kappa}_2$ goes to zero in the case of a constant magnetic field, and $\bar{\kappa}_1$ tends to zero in the case of a vanishing magnetic field. We also note that $\pi^{0}$ does not participate in the Hall-like thermal response of the medium, as they are electrically neutral and are not affected by the magnetic field.

\begin{figure*}
    \centering
    \centering
    \hspace{-2.5cm}
    \includegraphics[width=0.405\textwidth]{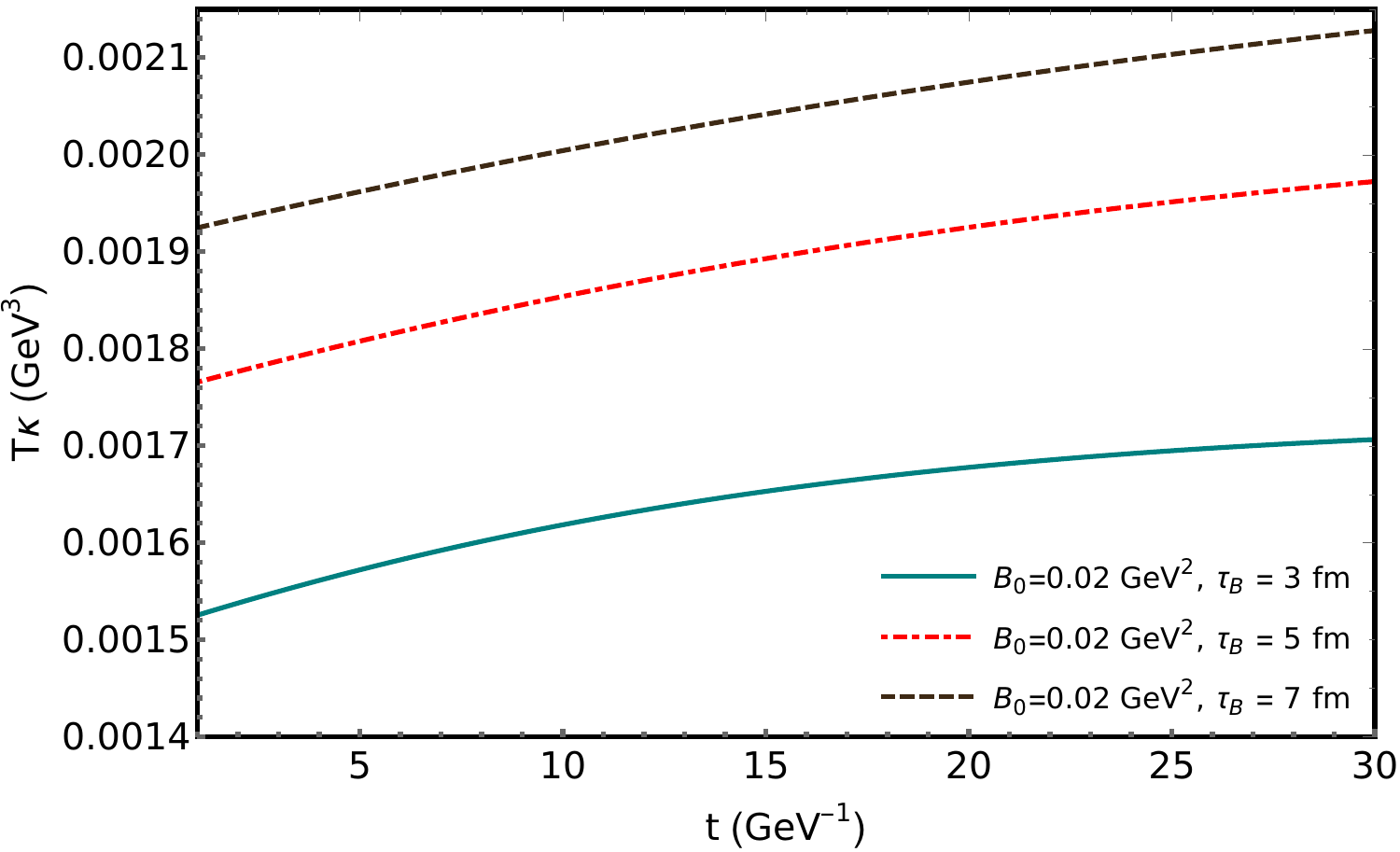}
    \hspace{0.2cm}
    \includegraphics[width=0.405\textwidth]{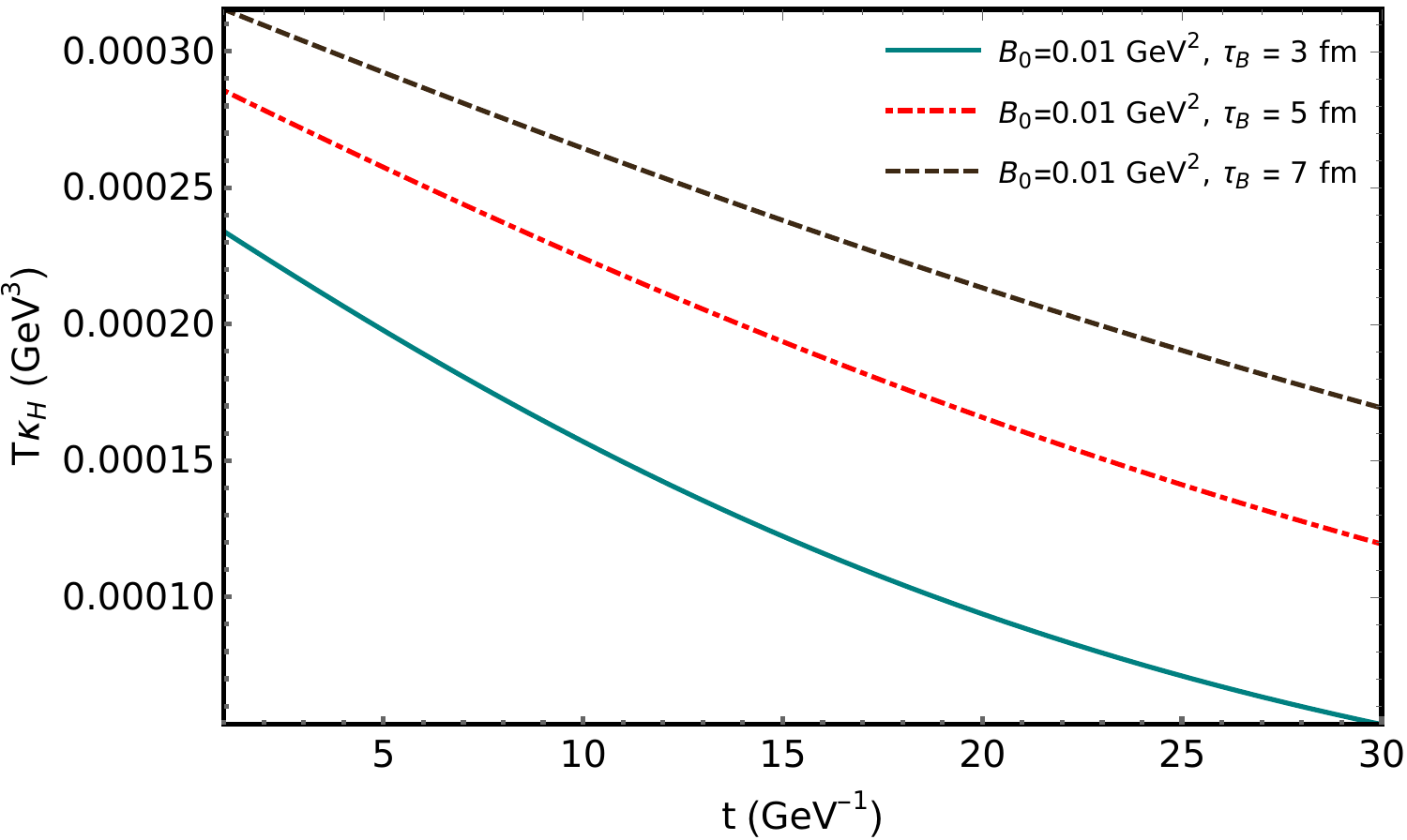}
    \hspace{-2.5cm}
    \caption{\small Thermal conductivity (left panel) and Hall-like Thermal conductivity (right panel) as a function of time for different values of decay parameter of the magnetic field at $T=0.14\,\text{GeV}$, $(\mu)_{\pi^{\pm}}=\pm0.1\,\text{GeV}$ and $(\mu)_{\pi^{0}}=0\,\text{GeV}$.  }
\label{f5}
\end{figure*}
\subsection*{Thermoelectric effect}
In the presence of a time-varying magnetic field, there are different sources of the induced electric field in the conducting hot pion gas. Most of the analyses consider the electric field due to the time decay of the magnetic field by using Faraday's law. Recently, it has been realized that the Seebeck effect can act as another source of the induced electric field, and is due to the local temperature gradient in the medium\cite{Bhatt:2018ncr}. Hence, the electric field induced in the medium ${\bf E}_{ind}$ can be expressed as,
{$${\bf E}_{ind} = {\bf E}_F+{\bf E}_T,$$
where ${\bf E}_F$ is the electric field due to Faraday's law, $\pmb{\nabla} \times \textbf{E} = \frac{\partial \textbf{B}}{\partial t}$ and ${\bf E}_T$ is the electric field induced due to the local temperature gradient in the medium}. 
\begin{figure*}
    \centering
    \centering
    \hspace{-2.5cm}
    \includegraphics[width=0.405\textwidth]{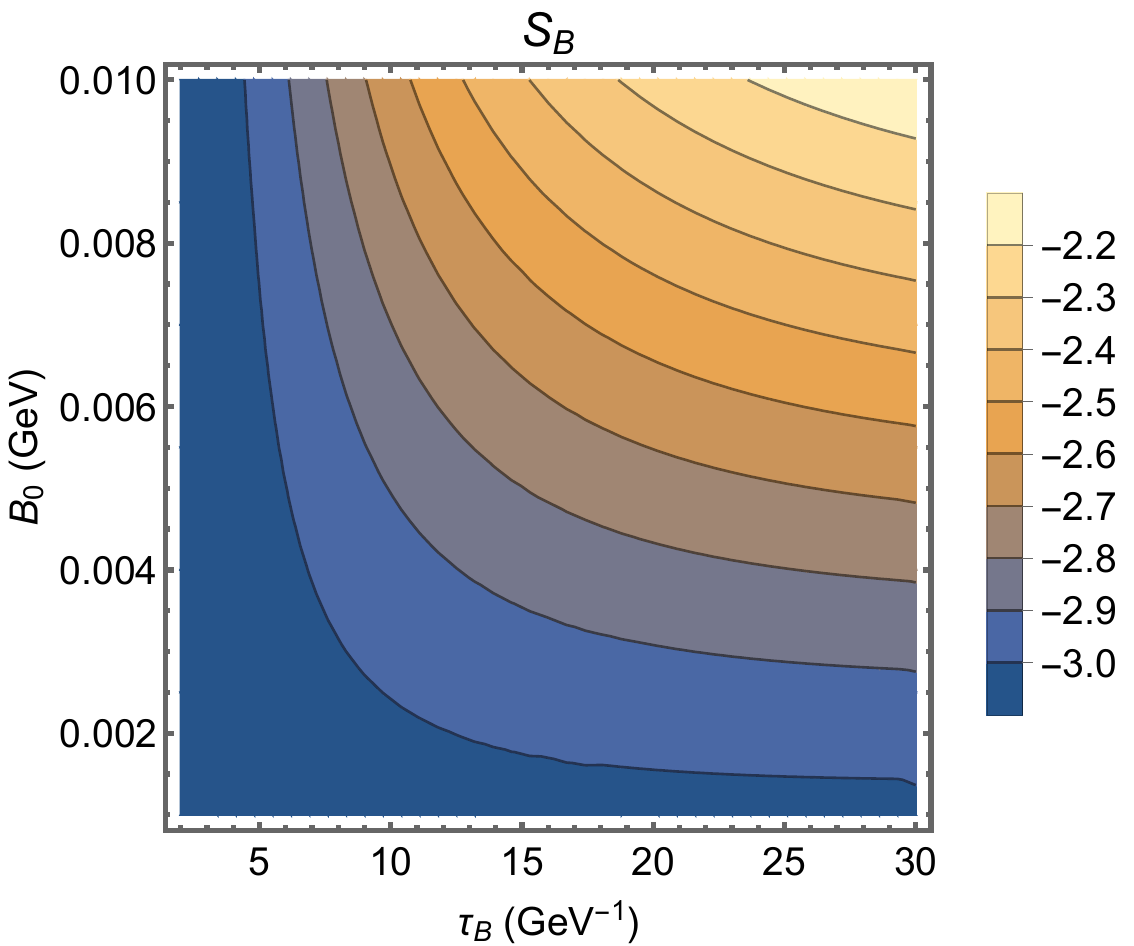}
    \hspace{0.2cm}
    \includegraphics[width=0.405\textwidth]{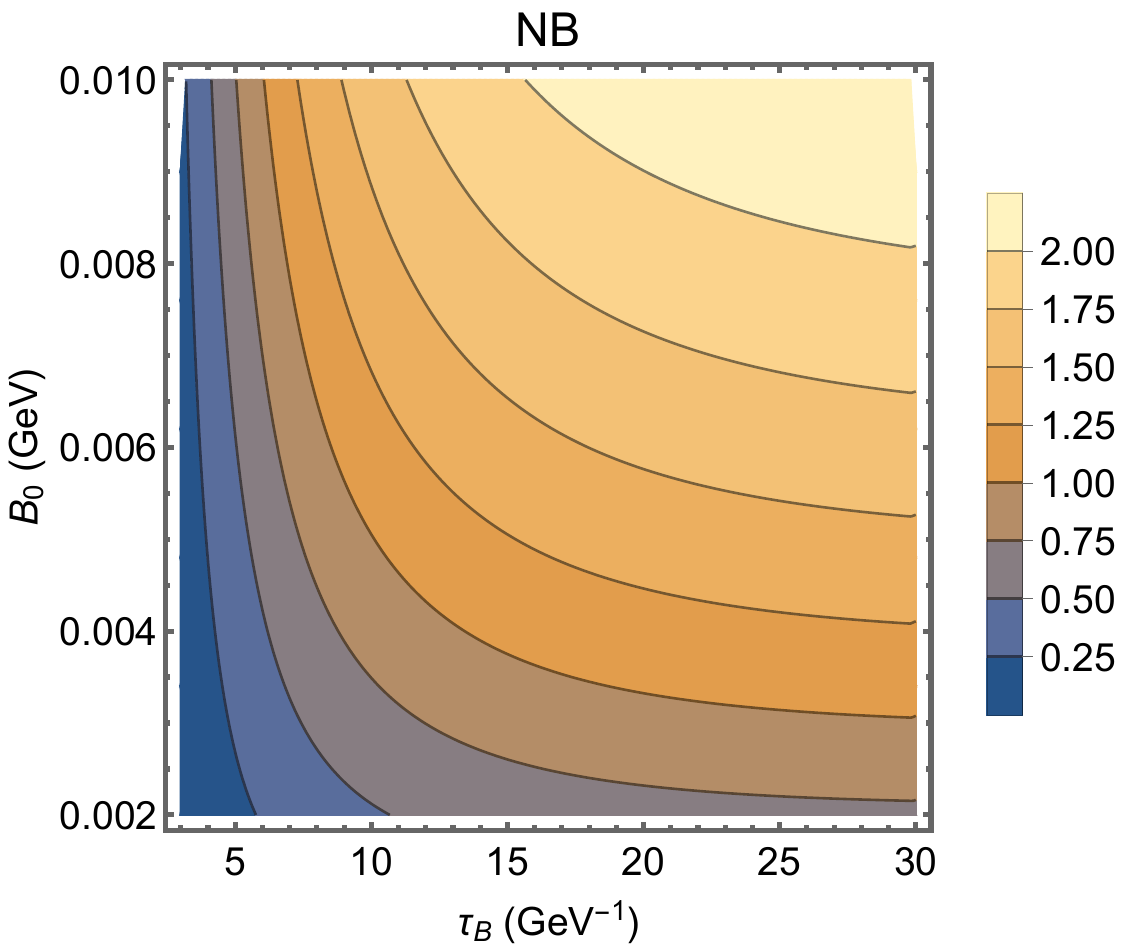}
    \hspace{-2.5cm}
    \caption{\small Magneto-Seebeck coefficient (left panel) and normalized Nernst coefficient (right panel) as a function of amplitude and decay parameter of the magnetic field at $t=2.5\,\text{fm}$, $T=0.12\,\text{GeV}$, $(\mu)_{\pi^{\pm}}=\pm0.1\,\text{GeV}$ and $(\mu)_{\pi^{0}}=0\,\text{GeV}$.}
\label{f6}
\end{figure*}
We follow the same prescription as in Ref.~\cite{PhysRevD.102.014030} to explore the thermoelectric effect of the hot pion gas. Here, we consider the case of an induced electric field from the temperature gradient, and hence we employ ${\bf E}_T\equiv {\bf E}$ with $E = |{\bf E}|$ in the rest of the analysis. The net current density of the hot pion gas is given by Eq.~\eqref{1}, 
with the non-equilibrium part of the distribution function $\delta f_k= f_k-f^0_k$ taking the form 
\begin{align}\label{IV.3}
  \delta f_k = \textbf{p}.[\gamma_1 \textbf{E} +\gamma_2\textbf{B}+ \gamma_3\textbf{X}+ \gamma_4(\textbf{X}\times \textbf{B})+ \gamma_5 \dot{\textbf{B}}\\ \nonumber +\gamma_6(\textbf{X}\times \dot{\textbf{B}}) + \gamma_7(\textbf{E}\times \textbf{B})+ \gamma_8(\textbf{E}\times \dot{\textbf{B}})]\frac{\partial f^0_k}{\partial \epsilon_k}.  
\end{align}
We now follow the same steps as before, i.e. we substitute Eq.\eqref{IV.3} in the Boltzmann equation \eqref{2}, obtain the coupled equations by comparing the tensorial structures on both sides, and solve them to obtain the $\gamma_i$ while considering $\textbf{B}.\textbf{X}=0$. We obtain
\begin{align}
 &\gamma_1 =-\frac{\tilde{\Omega}_k}{2}(I_1 e^{\eta_1} +I_2 e^{\eta_2} ),\\
 &\gamma_2 =0,\\ 
 &\gamma_3 =(\epsilon_k-h)\frac{F}{2 \epsilon_k} I_1 e^{\eta_1}+(\epsilon_k-h)\frac{F}{2 \epsilon_k} I_2 e^{\eta_1} ,\label{20.2}\\
 &\gamma_4 =i(\epsilon_k-h)\frac{ I_1}{2 \epsilon_k} e^{\eta_1} +i(\epsilon_k-h)\frac{I_2}{2 \epsilon_k} e^{\eta_1},\label{20.4}\\
 &\gamma_5 = 0,\label{20.1}\\
 &\gamma_6 =-\tau_R \gamma_4,\\
 &\gamma_7 =\frac{q_{k}i}{2\epsilon_k}(I_1 e^{\eta_1} -I_2 e^{\eta_2} ),\label{20.2}\\
 &\gamma_8 =-\frac{\tau_R q_{k}i}{2\epsilon_k}(I_1 e^{\eta_1} -I_2 e^{\eta_2}  ),\label{20.4}
\end{align} 
with, $\Tilde{\Omega}_k=\frac{q_{k}F}{\epsilon_k}$. Once we obtain the expression for $j$, we project it in two directions, along the direction of $\pmb{\nabla} T$ and $\pmb{\nabla} T \times \textbf{B}$, which we call directions $x$ and $y$, respectively. Then, the induced electric field can be obtained by considering the steady-state solution ($j_x,j_y =0$). This gives two coupled equations relating the induced electric fields ($E_x$, $E_y$) and the temperature gradients ($\frac{dT}{dx}$, $\frac{dT}{dy}$) along $x$ and $y$ directions. We can write them in a matrix form as follows
\begin{align}
 \begin{pmatrix}
E_x\\E_y
\end{pmatrix}
=
 \begin{pmatrix}
S_B & NB\\-NB &S_B
\end{pmatrix}
\begin{pmatrix}
 \frac{dT}{dx}\\ \frac{dT}{dy}
\end{pmatrix}.
\end{align}
where $S_B$ and $NB$ are the magneto-Seebeck and normalized Nernst coefficients, respectively. For a medium, $S_B$ and $NB$ characterize its thermoelectric behaviour, and in the presence of a time-varying magnetic field, they take the following forms 
\begin{align}
    &S_B = -\frac{L_1 L_4 +L_2 L_5+L_3 L_5 +L_2 L_6 +L_3 L_6 }{T(L_1^2+L_2^2+L_3^2 +2L_2L_3)},\\
    &NB = \frac{L_2 L_4 + L_3 L_4 -L_1 L_5 -L_1 L_6}{T(L_1^2+L_2^2+L_3^2 +2L_2L_3)},
\end{align}
where the integrals $L_i, i=1,2..6$ take the following forms, 
\begin{align}
    &L_1 = \frac{e}{3} \int dP\,p^2\bigg\{\big(\gamma_1\frac{\partial f^0}{\partial \epsilon}\big)_{\pi^+}-\big(\gamma_1\frac{\partial f^0}{\partial \epsilon}\big)_{\pi^-}\bigg\},\\
    &L_2 = \frac{Be}{3} \int dP\,p^2\bigg\{\big(\gamma_7\frac{\partial f^0}{\partial \epsilon}\big)_{\pi^+}-\big(\gamma_7\frac{\partial f^0}{\partial \epsilon}\big)_{\pi^-}\bigg\},\\
    &L_3 = \frac{\dot{B}e}{3} \int dP\,p^2\bigg\{\big(\gamma_8\frac{\partial f^0}{\partial \epsilon}\big)_{\pi^+}-\big(\gamma_8\frac{\partial f^0}{\partial \epsilon}\big)_{\pi^-}\bigg\}, \\
    &L_4 = \frac{e}{3} \int dP\,p^2\bigg\{\big(\gamma_3\frac{\partial f^0}{\partial \epsilon}\big)_{\pi^+}-\big(\gamma_3\frac{\partial f^0}{\partial \epsilon}\big)_{\pi^-}\bigg\},\\
    &L_5 = \frac{Be}{3} \int dP\,p^2\bigg\{\big(\gamma_4\frac{\partial f^0}{\partial \epsilon}\big)_{\pi^+}-\big(\gamma_4\frac{\partial f^0}{\partial \epsilon}\big)_{\pi^-}\bigg\},\\
    &L_6 = \frac{\dot{B}e}{3} \int dP\,p^2\bigg\{\big(\gamma_6\frac{\partial f^0}{\partial \epsilon}\big)_{\pi^+}-\big(\gamma_6\frac{\partial f^0}{\partial \epsilon}\big)_{\pi^-}\bigg\}.
\end{align}
We discuss the impact of the magnetic field, its time evolution, on the thermal and thermoelectric behavior of the hot pion gas in section \ref{V}. 

\section{Phenomenologically significant quantities}\label{IV}
In this section, we consider the significance of thermal transport in the presence of a time-evolving magnetic field in the context of HIC experiments. The thermal conductivity obtained can be employed to study the Knudsen number~\cite{rath2023effects}. The impact of thermal conductivity of the medium on the elliptic flow is explored in Ref.~\cite{Bhalerao:2005mm} for the case of a vanishing magnetic field. The current study focuses on the dependence of magnetic field evolution on thermal transport and its impact on the elliptic flow coefficient.
\begin{figure*}
    \centering
    \centering
    \hspace{-2.5cm}
    \includegraphics[width=0.405\textwidth]{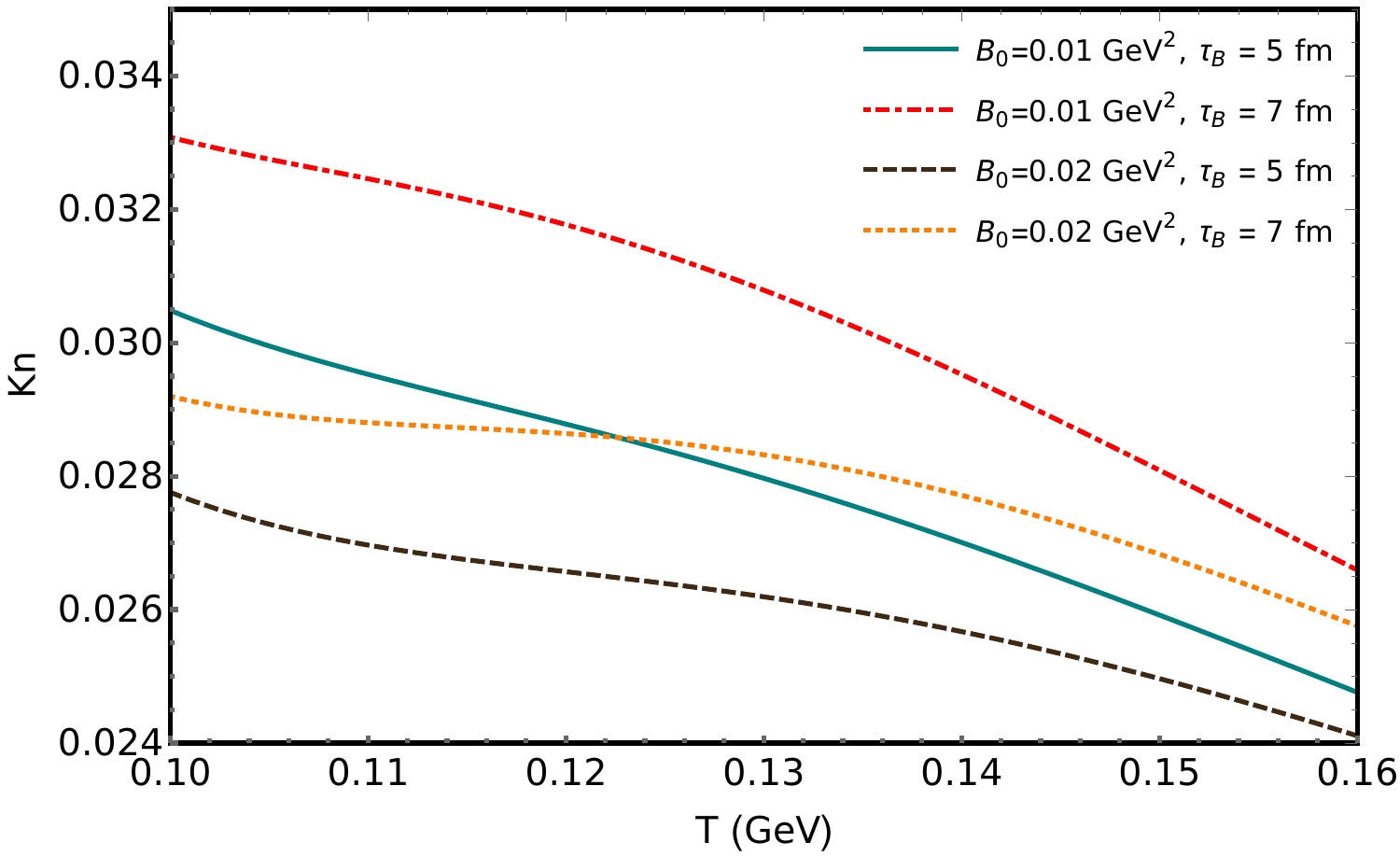}
    \hspace{0.2cm}
    \includegraphics[width=0.405\textwidth]{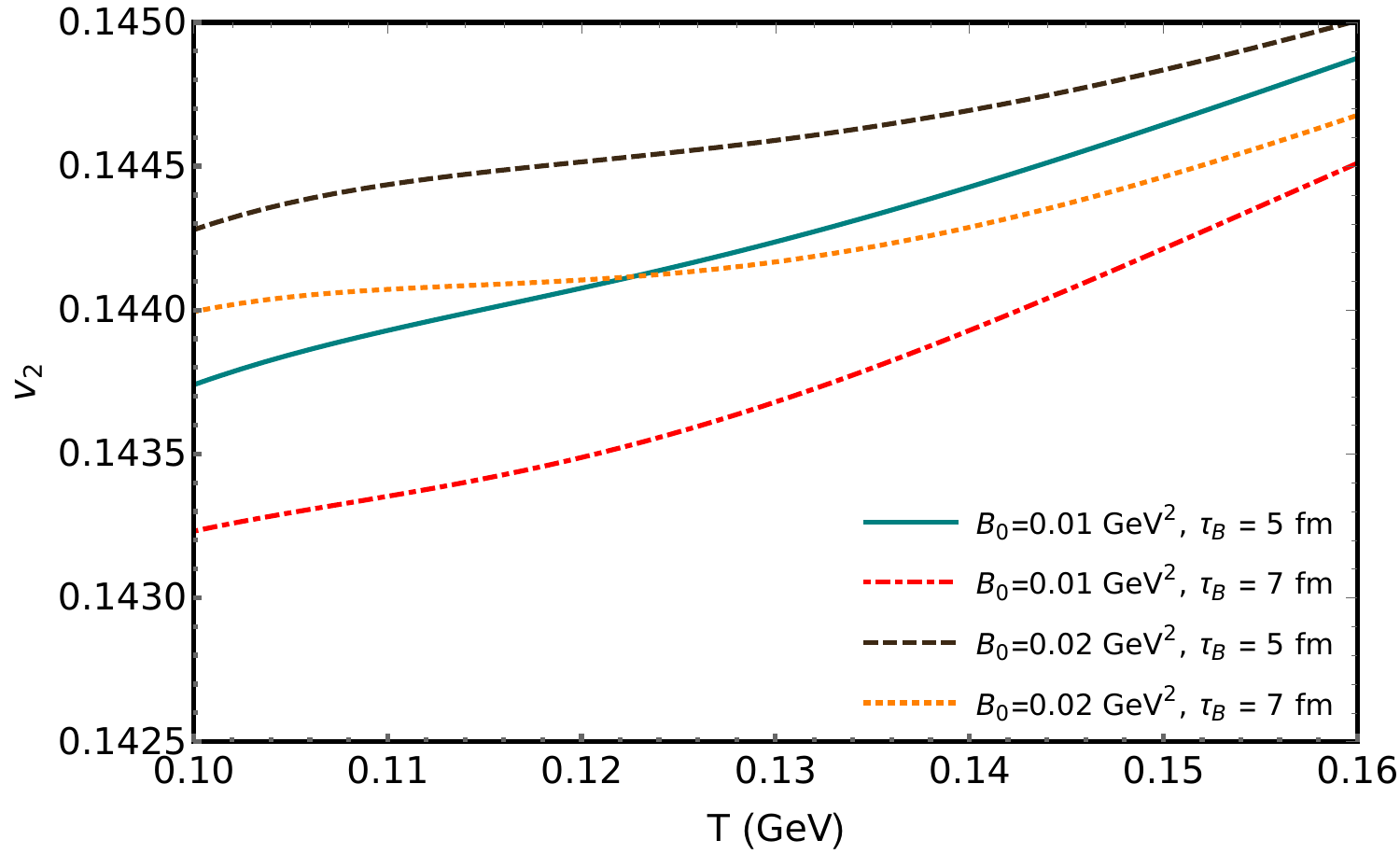}
    \hspace{-2.5cm}
    \caption{\small Knudsen number(left panel) and $v_2$ (right panel) as a function of temperature for different values of amplitude and decay parameter of the magnetic field at $t=2.5\,\text{fm}$, $(\mu)_{\pi^{\pm}}=\pm0.1\,\text{GeV}$ and $(\mu)_{\pi^{0}}=0\,\text{GeV}$.  }
\label{f7}
\end{figure*}

Knudsen number, $Kn$, is defined as the ratio of the mean path ($\lambda$) of the constituent particle to the size of the system, $l$,
\begin{align}\label{}
    Kn=\frac{\lambda}{l}.
\end{align}
If $Kn$ is equal to or greater than one, then the mean free path is comparable to the system size, and the continuum assumption of hydrodynamics is no longer applicable. Hydrodynamical modeling applies when the mean free path is much smaller than the characteristic system size, i.e. $Kn << 1$. The mean free path is related to the thermal conductivity as $\lambda = \frac{3\kappa_0}{vC_v}$
where $v$ is the relative speed, and $C_v$ is the specific heat at constant volume. Hence, the Knudsen number can be expressed in terms of thermal conductivity as, 
\begin{align}\label{56}
    Kn = \frac{3\kappa_0}{lvC_v}.
\end{align}
For the quantitative estimation, we have chosen $v \approx 0.98$, $l = 4$ fm~\cite{rath2022momentum}, and for $C_v$ we use the expression from \cite{PhysRevC.83.014906} and solve for the case of pion gas.

The elliptic flow $v_2$ can be expressed in terms of the Knudsen number as~\cite{Bhalerao:2005mm}, 
\begin{align}\label{22}
    v_2 = \frac{v_2^h}{1+\frac{Kn}{Kn_0}},
\end{align}
where $v_2^h$ is the elliptic flow at the hydrodynamical limit, $Kn \rightarrow 0$, the quantity $Kn_0$ is a number obtained to fit the Monte-Carlo simulations of the relativistic Boltzmann equation~\cite{Bhalerao:2005mm}. In our present analysis, we have taken $v_2^h =0.154 \pm 0.014$~\cite{PhysRevLett.91.182301} and $Kn_0 = 0.7$~\cite{Gombeaud:2007ub}. The effects of a time-varying magnetic field on the thermal conductivity and, hence, the elliptic flow are discussed in detail in section \ref{V}.

\section{Results and discussion}\label{V}
We start our discussion with the electrical response of the pion gas medium in the presence of time-dependent electromagnetic fields. This is measured using the electrical conductivity of the medium. The Ohmic and Hall conductivities are given as $\sigma_e=\frac{j_e^{(0)}+j_e^{(1)}}{E}$ and $\sigma_H=\frac{j_H^{(0)}+j_H^{(1)}+j_H^{(2)}}{E}$ respectively. Fig.~\ref{f1} shows the behaviour of $\sigma_e/T$ (left panel) and $\sigma_H/T$ (right panel) with temperature for different amplitudes of the magnetic field, respectively. The Electrical conductivity is seen to decrease with increasing temperature, as thermal effects become dominant and collisions take over the system's behavior. An increase in the magnetic field reduces the Ohmic conductivity as the charged particles are deflected in the direction perpendicular to the electric and magnetic fields, thereby enhancing the Hall conductivity. The Ohmic conductivity results show the same behavior and range as that of the results in Ref.~\cite{PhysRevD.102.076007,kalikotay2024electrical}. With increasing temperature, the influence of electromagnetic fields diminishes due to enhanced collisional effects, leading to a decrease in the Hall conductivity. In contrast, Hall conductivity increases with stronger magnetic fields, as more charged particles are deflected by the Lorentz force. The behavior and the range of Hall conductivity are seen to be consistent with the results of Ref.~\cite{PhysRevD.102.076007,kalikotay2024electrical}. 

Fig.~\ref{f2} shows the time behavior of Ohmic conductivity (left panel) and the Hall conductivity (right panel) of the pion gas. The ohmic current is seen to increase with time and saturate to a value at large times. This is because the magnetic field under consideration is an exponentially decaying one, and hence at large time goes to zero. This ensures the Ohmic current asymptotically approaches its maximum value at large time. An increase in the decay parameter of the fields is shown to delay the time it takes to reach its maximum value as the fields decay more slowly and stay relevant for longer. For a fixed decay rate, increasing the amplitude of the magnetic field reduces the Ohmic conductivity. The Hall conductivity decreases with time and vanishes at large values of time, as it is directly related to the magnetic field strength. A slower decay of the magnetic field (larger $\tau_B$) is shown to delay the decay of the Hall conductivity. An increase in the amplitude of the magnetic field for a fixed decay rate enhances the Hall conductivity. Both Ohmic and Hall conductivities show the behavior of crossover for the different decay rates. This crossover and the point at which it happens depend crucially on the temperature and occur due to the interplay between the decay of the fields and the interactions between the particles of the medium, controlled by the relaxation time of the medium. This can be observed through the following equation,   
\begin{equation}\label{tb}
    \eta_j = -\frac{t}{\tau_R} +a_j\frac{q_{k} i}{\epsilon_k}\int F dt.
\end{equation}
Here, $F$ contains the time behavior of the system. If we fix the temperature and look at the time behavior of the Ohmic and Hall conductivities, which of the two terms dominates in Eq.~\eqref{tb} depends on the time under consideration and the magnetic field. This complex interplay within the system is captured in Fig.~\ref{f2}. 

The effect of decay rate and the amplitude of the magnetic field on the Ohmic and Hall conductivities is seen through the contour graphs of Fig.~\ref{f3}. Fixing the time and the temperature of the system, we see that the Ohmic conductivity increases with the increase in the decay parameter and is affected minimally by the change in $B_0$ at low values of $\tau_B$. Hall conductivity, on the other hand, varies equally with both $B_0$ and $\tau_B$, except in the lower value region of $\tau_B$, where it seems to be unaffected by both. 

In Fig.~\ref{f4} we study the temperature dependence of the thermal conductivity (left panel) and Hall-like thermal conductivity ($\kappa_H=\bar{\kappa}_1+\bar{\kappa}_2$) (right panel), for different values of $B_0$ and $\tau_B$. The thermal conductivity is observed to increase with temperature and decrease with the increase in $B_0$. A faster decay rate of the field leads to a reduction in thermal conductivity. At high temperatures, the results corresponding to different values of $B_0$ converge for a fixed decay rate. The results are compared with the results of Ref.~\cite{PhysRevD.89.054013} and agree on the range and behavior of the thermal conductivities. The Hall-like thermal conductivity exhibits a strong dependence on both $B_0$ and $\tau_B$. Since $\kappa_H$ arises purely due to the presence of a magnetic field, it vanishes as the field approaches zero. Consequently, a more rapidly decaying magnetic field leads to a significant reduction in $\kappa_H$. A similar decline is observed with decreasing $B_0$. 

Fig.~\ref{f5} illustrates the time evolution of $\kappa$ (left panel) and $\kappa_H$ (right panel). It can be noted that the thermal conductivity of the medium increases with time as the strength of the magnetic field decreases. Also, for a given time, $\kappa$ increases as the decay rate of the field decreases. In contrast, the Hall-like thermal conductivity exhibits the opposite trend with time. As the magnetic field decays with time, $\kappa_H$ is seen to decrease. A magnetic field that persists for longer, i.e., has a larger $\tau_B$, supports $\kappa_H$ for longer. 

The contour plots for magneto-Seebeck and normalized Nernst coefficients with the amplitude of the magnetic field and decay parameter of the field are plotted in the left and right panels of Fig.~\ref{f6}, respectively. The magnitude of the magneto-Seebeck coefficient $S_B$ is seen to decrease with an increase in $B_0$ and $\tau_B$, showing comparable sensitivity to both. It responds slightly more to the change in $B_0$ and $\tau_B$ at higher values of $B_0$ and lower values of $\tau_B$. The decrease in $S_B$ with $B_0$ suggests that the induced electric field in the medium decreases with the increasing strength of the magnetic field. The Nernst coefficient, $NB$, increases with an increase in the magnetic field amplitude and a decrease in the decay rates. Higher decay rates are observed to have a stronger impact on the Nernst coefficient. The increase in $NB$ with $B_0$ is due to the fact that the electric field generated perpendicular to the temperature gradient and the magnetic field increases with the strength of the magnetic field.  

In Fig.~\ref{f7}, we plot the phenomenologically significant quantities, the Knudsen number, $Kn$, and the elliptic flow, $v_2$, with temperature on the left and right panels, respectively. The Knudsen number is found to be much smaller than unity, $Kn<<1$, indicating that the medium is dilute enough for hydrodynamics to be applicable. $Kn$ decreases with an increase in temperature, which is expected because an increase in temperature causes the mean free path to decrease. We also note that for a fixed temperature and decay rate of the magnetic field, $Kn$ decreases with an increase in $B_0$, and for a fixed temperature and $B_0$, it again decreases with an increase in the decay rate of the field. At high temperatures, the effect of $B_0$ diminishes, and is evident by the convergence of the lines representing different values of $B_0$ and the same values of $\tau_B$. The elliptic flow's behavior with temperature for different values of $B_0$ and $\tau_B$ is plotted in the right panel of Fig.~\ref{f7}. As anticipated from their relation in Eq.~\eqref{22}, $v_2$ exhibits a trend opposite to that of $Kn$  with respect to temperature, $B_0$, and $\tau_B$. The modeling indicates that both the amplitude and the decay rate of the magnetic field have a significant impact on $v_2$. An increase in $B_0$ increases the $v_2$ while an increase in $\tau_B$ decreases the elliptic flow. Hence, a highly conducting medium that decreases the decay rate of the magnetic field may create a more significant difference in $v_2$ as compared to central HICs with vanishing magnetic fields. A more precise understanding of this behavior would require a detailed investigation, which lies beyond the scope of the present work. 
\section{Conclusion and Outlook}\label{VI}
In this work, we have investigated the influence of time-dependent magnetic fields on the electrical, thermal and thermoelectric transport properties of a hot pion gas, and evaluated their impact on phenomenologically relevant observables in the context of heavy-ion collisions. This has been achieved by solving the relativistic Boltzmann equation within the relaxation time approximation, incorporating a temperature-dependent parametrization for the pion relaxation time.

We have found that both Ohmic and Hall conductivities exhibit explicit dependence on the temperature of the medium as well as on the amplitude and decay rate of the magnetic field. The time evolution of the electrical conductivities displays non-trivial behavior and is due to the complex interplay between temperature and magnetic field within the pionic system. The results are consistent with earlier studies in the constant-field limit, while extending the analysis to more realistic, time-evolving electromagnetic backgrounds.

The thermal response of the medium has been characterized by both the longitudinal thermal conductivity and the transverse Hall-like component, defined with respect to the thermal driving force. Time-dependent magnetic fields have been shown to influence these transport coefficients significantly via their decay dynamics. In addition, the thermoelectric response, quantified through the Seebeck and Nernst coefficients, demonstrates a strong sensitivity to the magnetic field profile, with both the field amplitude $B_0$ and the decay parameter, $\tau_{B}$, playing crucial roles.

Finally, the impact of these transport coefficients on phenomenologically significant quantities, such as the Knudsen number and elliptic flow coefficient, $v_2$ is analyzed. The results underscore that evolving magnetic fields introduce meaningful corrections to these observables and are important for characterizing the collective behavior of the medium. These findings highlight the importance of incorporating time-dependent electromagnetic field evolution into theoretical descriptions of the hadronic phase in heavy-ion collisions. Such considerations are essential for a more accurate understanding of the transport dynamics and their phenomenological implications, particularly in relation to observables accessible at RHIC and LHC.

Overall, our analysis demonstrates that time dependent magnetic fields have a substantial and non-negligible influence on the transport dynamics of the hadronic phase. These findings highlight the importance of incorporating realistic electromagnetic field evolution, in conjunction with accurate hadronic transport modeling, to achieve a comprehensive understanding of the late-time dynamics in heavy-ion collisions. The framework developed here offers a more realistic basis for interpreting experimental data from RHIC and LHC and paves the way for future studies involving more complex interactions and evolving medium profiles.

Incorporating a dynamically evolving medium and accounting for spatial gradients in temperature, flow velocity, and chemical potential would provide additional corrections to the transport coefficients and yield a more complete picture of the system’s evolution. The current treatment, based on a parametrized relaxation time within the RTA, can be refined by considering the magnetic field dependence of the relaxation time, as well as adopting more realistic, number-conserving collision integrals such as those in the energy-dependent relaxation time approximation (ERTA)~\cite{KumarSingh:2025kml, Singh:2024leo}.

Moreover, incorporating the calculated transport coefficients into full (3+1)D hydrodynamic or hybrid simulations would enable direct comparisons with experimental data, thereby offering a way to constrain the evolution and lifetime of magnetic fields in heavy-ion collisions. Such studies would be instrumental in bridging microscopic transport theory with final-state observables and in addressing the open questions surrounding electromagnetic field dynamics in the QCD medium created at RHIC and LHC.

\section*{Acknowledgments}
G.K.K. is thankful to the Indian Institute of Technology Bombay for the Institute postdoctoral fellowship. S. D. acknowledges the SERB
Power Fellowship, SPF/2022/000014 for the support on this work.
\bibliography{ref}{}

\begin{thebibliography}{50}%
\makeatletter
\providecommand \@ifxundefined [1]{%
 \@ifx{#1\undefined}
}%
\providecommand \@ifnum [1]{%
 \ifnum #1\expandafter \@firstoftwo
 \else \expandafter \@secondoftwo
 \fi
}%
\providecommand \@ifx [1]{%
 \ifx #1\expandafter \@firstoftwo
 \else \expandafter \@secondoftwo
 \fi
}%
\providecommand \natexlab [1]{#1}%
\providecommand \enquote  [1]{``#1''}%
\providecommand \bibnamefont  [1]{#1}%
\providecommand \bibfnamefont [1]{#1}%
\providecommand \citenamefont [1]{#1}%
\providecommand \href@noop [0]{\@secondoftwo}%
\providecommand \href [0]{\begingroup \@sanitize@url \@href}%
\providecommand \@href[1]{\@@startlink{#1}\@@href}%
\providecommand \@@href[1]{\endgroup#1\@@endlink}%
\providecommand \@sanitize@url [0]{\catcode `\\12\catcode `\$12\catcode `\&12\catcode `\#12\catcode `\^12\catcode `\_12\catcode `\%12\relax}%
\providecommand \@@startlink[1]{}%
\providecommand \@@endlink[0]{}%
\providecommand \url  [0]{\begingroup\@sanitize@url \@url }%
\providecommand \@url [1]{\endgroup\@href {#1}{\urlprefix }}%
\providecommand \urlprefix  [0]{URL }%
\providecommand \Eprint [0]{\href }%
\providecommand \doibase [0]{https://doi.org/}%
\providecommand \selectlanguage [0]{\@gobble}%
\providecommand \bibinfo  [0]{\@secondoftwo}%
\providecommand \bibfield  [0]{\@secondoftwo}%
\providecommand \translation [1]{[#1]}%
\providecommand \BibitemOpen [0]{}%
\providecommand \bibitemStop [0]{}%
\providecommand \bibitemNoStop [0]{.\EOS\space}%
\providecommand \EOS [0]{\spacefactor3000\relax}%
\providecommand \BibitemShut  [1]{\csname bibitem#1\endcsname}%
\let\auto@bib@innerbib\@empty
\bibitem [{\citenamefont {Adams}\ \emph {et~al.}(2005)\citenamefont {Adams} \emph {et~al.}}]{Adams:2005dq}%
  \BibitemOpen
  \bibfield  {author} {\bibinfo {author} {\bibfnamefont {J.}~\bibnamefont {Adams}} \emph {et~al.} (\bibinfo {collaboration} {STAR}),\ }\bibfield  {title} {\bibinfo {title} {{Experimental and theoretical challenges in the search for the quark gluon plasma: The STAR Collaboration's critical assessment of the evidence from RHIC collisions}},\ }\href {https://doi.org/10.1016/j.nuclphysa.2005.03.085} {\bibfield  {journal} {\bibinfo  {journal} {Nucl. Phys. A}\ }\textbf {\bibinfo {volume} {757}},\ \bibinfo {pages} {102} (\bibinfo {year} {2005})},\ \Eprint {https://arxiv.org/abs/nucl-ex/0501009} {arXiv:nucl-ex/0501009} \BibitemShut {NoStop}%
\bibitem [{\citenamefont {Adcox}\ \emph {et~al.}(2005)\citenamefont {Adcox} \emph {et~al.}}]{Adcox:2004mh}%
  \BibitemOpen
  \bibfield  {author} {\bibinfo {author} {\bibfnamefont {K.}~\bibnamefont {Adcox}} \emph {et~al.} (\bibinfo {collaboration} {PHENIX}),\ }\bibfield  {title} {\bibinfo {title} {{Formation of dense partonic matter in relativistic nucleus-nucleus collisions at RHIC: Experimental evaluation by the PHENIX collaboration}},\ }\href {https://doi.org/10.1016/j.nuclphysa.2005.03.086} {\bibfield  {journal} {\bibinfo  {journal} {Nucl. Phys. A}\ }\textbf {\bibinfo {volume} {757}},\ \bibinfo {pages} {184} (\bibinfo {year} {2005})},\ \Eprint {https://arxiv.org/abs/nucl-ex/0410003} {arXiv:nucl-ex/0410003} \BibitemShut {NoStop}%
\bibitem [{\citenamefont {Back}\ \emph {et~al.}(2005)\citenamefont {Back} \emph {et~al.}}]{Back:2004je}%
  \BibitemOpen
  \bibfield  {author} {\bibinfo {author} {\bibfnamefont {B.}~\bibnamefont {Back}} \emph {et~al.} (\bibinfo {collaboration} {PHOBOS}),\ }\bibfield  {title} {\bibinfo {title} {{The PHOBOS perspective on discoveries at RHIC}},\ }\href {https://doi.org/10.1016/j.nuclphysa.2005.03.084} {\bibfield  {journal} {\bibinfo  {journal} {Nucl. Phys. A}\ }\textbf {\bibinfo {volume} {757}},\ \bibinfo {pages} {28} (\bibinfo {year} {2005})},\ \Eprint {https://arxiv.org/abs/nucl-ex/0410022} {arXiv:nucl-ex/0410022} \BibitemShut {NoStop}%
\bibitem [{\citenamefont {Arsene}\ \emph {et~al.}(2005)\citenamefont {Arsene} \emph {et~al.}}]{Arsene:2004fa}%
  \BibitemOpen
  \bibfield  {author} {\bibinfo {author} {\bibfnamefont {I.}~\bibnamefont {Arsene}} \emph {et~al.} (\bibinfo {collaboration} {BRAHMS}),\ }\bibfield  {title} {\bibinfo {title} {{Quark gluon plasma and color glass condensate at RHIC? The Perspective from the BRAHMS experiment}},\ }\href {https://doi.org/10.1016/j.nuclphysa.2005.02.130} {\bibfield  {journal} {\bibinfo  {journal} {Nucl. Phys. A}\ }\textbf {\bibinfo {volume} {757}},\ \bibinfo {pages} {1} (\bibinfo {year} {2005})},\ \Eprint {https://arxiv.org/abs/nucl-ex/0410020} {arXiv:nucl-ex/0410020} \BibitemShut {NoStop}%
\bibitem [{\citenamefont {Aamodt}\ \emph {et~al.}(2010)\citenamefont {Aamodt} \emph {et~al.}}]{Aamodt:2010pb}%
  \BibitemOpen
  \bibfield  {author} {\bibinfo {author} {\bibfnamefont {K.}~\bibnamefont {Aamodt}} \emph {et~al.} (\bibinfo {collaboration} {ALICE}),\ }\bibfield  {title} {\bibinfo {title} {{Charged-particle multiplicity density at mid-rapidity in central Pb-Pb collisions at $\sqrt{s_{NN}} = 2.76$ TeV}},\ }\href {https://doi.org/10.1103/PhysRevLett.105.252301} {\bibfield  {journal} {\bibinfo  {journal} {Phys. Rev. Lett.}\ }\textbf {\bibinfo {volume} {105}},\ \bibinfo {pages} {252301} (\bibinfo {year} {2010})},\ \Eprint {https://arxiv.org/abs/1011.3916} {arXiv:1011.3916 [nucl-ex]} \BibitemShut {NoStop}%
\bibitem [{\citenamefont {Turbide}\ \emph {et~al.}(2004)\citenamefont {Turbide}, \citenamefont {Rapp},\ and\ \citenamefont {Gale}}]{PhysRevC.69.014903}%
  \BibitemOpen
  \bibfield  {author} {\bibinfo {author} {\bibfnamefont {S.}~\bibnamefont {Turbide}}, \bibinfo {author} {\bibfnamefont {R.}~\bibnamefont {Rapp}},\ and\ \bibinfo {author} {\bibfnamefont {C.}~\bibnamefont {Gale}},\ }\bibfield  {title} {\bibinfo {title} {Hadronic production of thermal photons},\ }\href {https://doi.org/10.1103/PhysRevC.69.014903} {\bibfield  {journal} {\bibinfo  {journal} {Phys. Rev. C}\ }\textbf {\bibinfo {volume} {69}},\ \bibinfo {pages} {014903} (\bibinfo {year} {2004})}\BibitemShut {NoStop}%
\bibitem [{\citenamefont {Steele}\ \emph {et~al.}(1996)\citenamefont {Steele}, \citenamefont {Yamagishi},\ and\ \citenamefont {Zahed}}]{STEELE1996255}%
  \BibitemOpen
  \bibfield  {author} {\bibinfo {author} {\bibfnamefont {J.~V.}\ \bibnamefont {Steele}}, \bibinfo {author} {\bibfnamefont {H.}~\bibnamefont {Yamagishi}},\ and\ \bibinfo {author} {\bibfnamefont {I.}~\bibnamefont {Zahed}},\ }\bibfield  {title} {\bibinfo {title} {Dilepton and photon emission rates from a hadronic gas},\ }\href {https://doi.org/https://doi.org/10.1016/0370-2693(96)00802-7} {\bibfield  {journal} {\bibinfo  {journal} {Physics Letters B}\ }\textbf {\bibinfo {volume} {384}},\ \bibinfo {pages} {255} (\bibinfo {year} {1996})}\BibitemShut {NoStop}%
\bibitem [{\citenamefont {Dash}\ \emph {et~al.}(2020)\citenamefont {Dash}, \citenamefont {Samanta}, \citenamefont {Dey}, \citenamefont {Gangopadhyaya}, \citenamefont {Ghosh},\ and\ \citenamefont {Roy}}]{Dash:2020vxk}%
  \BibitemOpen
  \bibfield  {author} {\bibinfo {author} {\bibfnamefont {A.}~\bibnamefont {Dash}}, \bibinfo {author} {\bibfnamefont {S.}~\bibnamefont {Samanta}}, \bibinfo {author} {\bibfnamefont {J.}~\bibnamefont {Dey}}, \bibinfo {author} {\bibfnamefont {U.}~\bibnamefont {Gangopadhyaya}}, \bibinfo {author} {\bibfnamefont {S.}~\bibnamefont {Ghosh}},\ and\ \bibinfo {author} {\bibfnamefont {V.}~\bibnamefont {Roy}},\ }\bibfield  {title} {\bibinfo {title} {{Anisotropic transport properties of a hadron resonance gas in a magnetic field}},\ }\href {https://doi.org/10.1103/PhysRevD.102.016016} {\bibfield  {journal} {\bibinfo  {journal} {Phys. Rev. D}\ }\textbf {\bibinfo {volume} {102}},\ \bibinfo {pages} {016016} (\bibinfo {year} {2020})},\ \Eprint {https://arxiv.org/abs/2002.08781} {arXiv:2002.08781 [nucl-th]} \BibitemShut {NoStop}%
\bibitem [{\citenamefont {Das}\ \emph {et~al.}(2019{\natexlab{a}})\citenamefont {Das}, \citenamefont {Mishra},\ and\ \citenamefont {Mohapatra}}]{Das:2019pqd}%
  \BibitemOpen
  \bibfield  {author} {\bibinfo {author} {\bibfnamefont {A.}~\bibnamefont {Das}}, \bibinfo {author} {\bibfnamefont {H.}~\bibnamefont {Mishra}},\ and\ \bibinfo {author} {\bibfnamefont {R.~K.}\ \bibnamefont {Mohapatra}},\ }\bibfield  {title} {\bibinfo {title} {{Transport coefficients of hot and dense hadron gas in a magnetic field: a relaxation time approach}},\ }\href {https://doi.org/10.1103/PhysRevD.100.114004} {\bibfield  {journal} {\bibinfo  {journal} {Phys. Rev. D}\ }\textbf {\bibinfo {volume} {100}},\ \bibinfo {pages} {114004} (\bibinfo {year} {2019}{\natexlab{a}})},\ \Eprint {https://arxiv.org/abs/1909.06202} {arXiv:1909.06202 [hep-ph]} \BibitemShut {NoStop}%
\bibitem [{\citenamefont {Kadam}\ \emph {et~al.}(2018)\citenamefont {Kadam}, \citenamefont {Mishra},\ and\ \citenamefont {Thakur}}]{Kadam:2017iaz}%
  \BibitemOpen
  \bibfield  {author} {\bibinfo {author} {\bibfnamefont {G.~P.}\ \bibnamefont {Kadam}}, \bibinfo {author} {\bibfnamefont {H.}~\bibnamefont {Mishra}},\ and\ \bibinfo {author} {\bibfnamefont {L.}~\bibnamefont {Thakur}},\ }\bibfield  {title} {\bibinfo {title} {{Electrical and thermal conductivities of hot and dense hadronic matter}},\ }\href {https://doi.org/10.1103/PhysRevD.98.114001} {\bibfield  {journal} {\bibinfo  {journal} {Phys. Rev. D}\ }\textbf {\bibinfo {volume} {98}},\ \bibinfo {pages} {114001} (\bibinfo {year} {2018})},\ \Eprint {https://arxiv.org/abs/1712.03805} {arXiv:1712.03805 [hep-ph]} \BibitemShut {NoStop}%
\bibitem [{\citenamefont {Kharzeev}\ \emph {et~al.}(2008)\citenamefont {Kharzeev}, \citenamefont {McLerran},\ and\ \citenamefont {Warringa}}]{KHARZEEV2008227}%
  \BibitemOpen
  \bibfield  {author} {\bibinfo {author} {\bibfnamefont {D.~E.}\ \bibnamefont {Kharzeev}}, \bibinfo {author} {\bibfnamefont {L.~D.}\ \bibnamefont {McLerran}},\ and\ \bibinfo {author} {\bibfnamefont {H.~J.}\ \bibnamefont {Warringa}},\ }\bibfield  {title} {\bibinfo {title} {The effects of topological charge change in heavy ion collisions: “event by event p and cp violation”},\ }\href {https://doi.org/https://doi.org/10.1016/j.nuclphysa.2008.02.298} {\bibfield  {journal} {\bibinfo  {journal} {Nuclear Physics A}\ }\textbf {\bibinfo {volume} {803}},\ \bibinfo {pages} {227} (\bibinfo {year} {2008})}\BibitemShut {NoStop}%
\bibitem [{\citenamefont {Skokov}\ \emph {et~al.}(2009)\citenamefont {Skokov}, \citenamefont {Illarionov},\ and\ \citenamefont {Toneev}}]{Skokov:2009qp}%
  \BibitemOpen
  \bibfield  {author} {\bibinfo {author} {\bibfnamefont {V.}~\bibnamefont {Skokov}}, \bibinfo {author} {\bibfnamefont {A.}~\bibnamefont {Illarionov}},\ and\ \bibinfo {author} {\bibfnamefont {V.}~\bibnamefont {Toneev}},\ }\bibfield  {title} {\bibinfo {title} {{Estimate of the magnetic field strength in heavy-ion collisions}},\ }\href {https://doi.org/10.1142/S0217751X09047570} {\bibfield  {journal} {\bibinfo  {journal} {Int. J. Mod. Phys. A}\ }\textbf {\bibinfo {volume} {24}},\ \bibinfo {pages} {5925} (\bibinfo {year} {2009})},\ \Eprint {https://arxiv.org/abs/0907.1396} {arXiv:0907.1396 [nucl-th]} \BibitemShut {NoStop}%
\bibitem [{\citenamefont {SKOKOV}\ \emph {et~al.}(2009)\citenamefont {SKOKOV}, \citenamefont {ILLARIONOV},\ and\ \citenamefont {TONEEV}}]{doi:10.1142/S0217751X09047570}%
  \BibitemOpen
  \bibfield  {author} {\bibinfo {author} {\bibfnamefont {V.~V.}\ \bibnamefont {SKOKOV}}, \bibinfo {author} {\bibfnamefont {A.~Y.}\ \bibnamefont {ILLARIONOV}},\ and\ \bibinfo {author} {\bibfnamefont {V.~D.}\ \bibnamefont {TONEEV}},\ }\bibfield  {title} {\bibinfo {title} {Estimate of the magnetic field strength in heavy-ion collisions},\ }\href {https://doi.org/10.1142/S0217751X09047570} {\bibfield  {journal} {\bibinfo  {journal} {International Journal of Modern Physics A}\ }\textbf {\bibinfo {volume} {24}},\ \bibinfo {pages} {5925} (\bibinfo {year} {2009})},\ \Eprint {https://arxiv.org/abs/https://doi.org/10.1142/S0217751X09047570} {https://doi.org/10.1142/S0217751X09047570} \BibitemShut {NoStop}%
\bibitem [{\citenamefont {Huang}\ \emph {et~al.}(2023)\citenamefont {Huang}, \citenamefont {She}, \citenamefont {Shi}, \citenamefont {Huang},\ and\ \citenamefont {Liao}}]{PhysRevC.107.034901}%
  \BibitemOpen
  \bibfield  {author} {\bibinfo {author} {\bibfnamefont {A.}~\bibnamefont {Huang}}, \bibinfo {author} {\bibfnamefont {D.}~\bibnamefont {She}}, \bibinfo {author} {\bibfnamefont {S.}~\bibnamefont {Shi}}, \bibinfo {author} {\bibfnamefont {M.}~\bibnamefont {Huang}},\ and\ \bibinfo {author} {\bibfnamefont {J.}~\bibnamefont {Liao}},\ }\bibfield  {title} {\bibinfo {title} {Dynamical magnetic fields in heavy-ion collisions},\ }\href {https://doi.org/10.1103/PhysRevC.107.034901} {\bibfield  {journal} {\bibinfo  {journal} {Phys. Rev. C}\ }\textbf {\bibinfo {volume} {107}},\ \bibinfo {pages} {034901} (\bibinfo {year} {2023})}\BibitemShut {NoStop}%
\bibitem [{\citenamefont {Inghirami}\ \emph {et~al.}(2020)\citenamefont {Inghirami}, \citenamefont {Mace}, \citenamefont {Hirono}, \citenamefont {Del~Zanna}, \citenamefont {Kharzeev},\ and\ \citenamefont {Bleicher}}]{inghirami2020magnetic}%
  \BibitemOpen
  \bibfield  {author} {\bibinfo {author} {\bibfnamefont {G.}~\bibnamefont {Inghirami}}, \bibinfo {author} {\bibfnamefont {M.}~\bibnamefont {Mace}}, \bibinfo {author} {\bibfnamefont {Y.}~\bibnamefont {Hirono}}, \bibinfo {author} {\bibfnamefont {L.}~\bibnamefont {Del~Zanna}}, \bibinfo {author} {\bibfnamefont {D.~E.}\ \bibnamefont {Kharzeev}},\ and\ \bibinfo {author} {\bibfnamefont {M.}~\bibnamefont {Bleicher}},\ }\bibfield  {title} {\bibinfo {title} {Magnetic fields in heavy ion collisions: flow and charge transport},\ }\href {https://doi.org/10.1140/epjc/s10052-020-7847-4} {\bibfield  {journal} {\bibinfo  {journal} {The European Physical Journal C}\ }\textbf {\bibinfo {volume} {80}},\ \bibinfo {pages} {293} (\bibinfo {year} {2020})}\BibitemShut {NoStop}%
\bibitem [{\citenamefont {Fukushima}\ \emph {et~al.}(2008)\citenamefont {Fukushima}, \citenamefont {Kharzeev},\ and\ \citenamefont {Warringa}}]{PhysRevD.78.074033}%
  \BibitemOpen
  \bibfield  {author} {\bibinfo {author} {\bibfnamefont {K.}~\bibnamefont {Fukushima}}, \bibinfo {author} {\bibfnamefont {D.~E.}\ \bibnamefont {Kharzeev}},\ and\ \bibinfo {author} {\bibfnamefont {H.~J.}\ \bibnamefont {Warringa}},\ }\bibfield  {title} {\bibinfo {title} {Chiral magnetic effect},\ }\href {https://doi.org/10.1103/PhysRevD.78.074033} {\bibfield  {journal} {\bibinfo  {journal} {Phys. Rev. D}\ }\textbf {\bibinfo {volume} {78}},\ \bibinfo {pages} {074033} (\bibinfo {year} {2008})}\BibitemShut {NoStop}%
\bibitem [{\citenamefont {Kharzeev}\ \emph {et~al.}(2016)\citenamefont {Kharzeev}, \citenamefont {Liao}, \citenamefont {Voloshin},\ and\ \citenamefont {Wang}}]{Kharzeev:2015znc}%
  \BibitemOpen
  \bibfield  {author} {\bibinfo {author} {\bibfnamefont {D.}~\bibnamefont {Kharzeev}}, \bibinfo {author} {\bibfnamefont {J.}~\bibnamefont {Liao}}, \bibinfo {author} {\bibfnamefont {S.}~\bibnamefont {Voloshin}},\ and\ \bibinfo {author} {\bibfnamefont {G.}~\bibnamefont {Wang}},\ }\bibfield  {title} {\bibinfo {title} {{Chiral magnetic and vortical effects in high-energy nuclear collisions\textemdash{}A status report}},\ }\href {https://doi.org/10.1016/j.ppnp.2016.01.001} {\bibfield  {journal} {\bibinfo  {journal} {Prog. Part. Nucl. Phys.}\ }\textbf {\bibinfo {volume} {88}},\ \bibinfo {pages} {1} (\bibinfo {year} {2016})},\ \Eprint {https://arxiv.org/abs/1511.04050} {arXiv:1511.04050 [hep-ph]} \BibitemShut {NoStop}%
\bibitem [{\citenamefont {Sadofyev}\ and\ \citenamefont {Isachenkov}(2011)}]{Sadofyev:2010pr}%
  \BibitemOpen
  \bibfield  {author} {\bibinfo {author} {\bibfnamefont {A.}~\bibnamefont {Sadofyev}}\ and\ \bibinfo {author} {\bibfnamefont {M.}~\bibnamefont {Isachenkov}},\ }\bibfield  {title} {\bibinfo {title} {{The Chiral magnetic effect in hydrodynamical approach}},\ }\href {https://doi.org/10.1016/j.physletb.2011.02.041} {\bibfield  {journal} {\bibinfo  {journal} {Phys. Lett. B}\ }\textbf {\bibinfo {volume} {697}},\ \bibinfo {pages} {404} (\bibinfo {year} {2011})},\ \Eprint {https://arxiv.org/abs/1010.1550} {arXiv:1010.1550 [hep-th]} \BibitemShut {NoStop}%
\bibitem [{\citenamefont {Kharzeev}\ and\ \citenamefont {Yee}(2011)}]{PhysRevD.83.085007}%
  \BibitemOpen
  \bibfield  {author} {\bibinfo {author} {\bibfnamefont {D.~E.}\ \bibnamefont {Kharzeev}}\ and\ \bibinfo {author} {\bibfnamefont {H.-U.}\ \bibnamefont {Yee}},\ }\bibfield  {title} {\bibinfo {title} {Chiral magnetic wave},\ }\href {https://doi.org/10.1103/PhysRevD.83.085007} {\bibfield  {journal} {\bibinfo  {journal} {Phys. Rev. D}\ }\textbf {\bibinfo {volume} {83}},\ \bibinfo {pages} {085007} (\bibinfo {year} {2011})}\BibitemShut {NoStop}%
\bibitem [{\citenamefont {Tuchin}(2013{\natexlab{a}})}]{PhysRevC.88.024910}%
  \BibitemOpen
  \bibfield  {author} {\bibinfo {author} {\bibfnamefont {K.}~\bibnamefont {Tuchin}},\ }\bibfield  {title} {\bibinfo {title} {Magnetic contribution to dilepton production in heavy-ion collisions},\ }\href {https://doi.org/10.1103/PhysRevC.88.024910} {\bibfield  {journal} {\bibinfo  {journal} {Phys. Rev. C}\ }\textbf {\bibinfo {volume} {88}},\ \bibinfo {pages} {024910} (\bibinfo {year} {2013}{\natexlab{a}})}\BibitemShut {NoStop}%
\bibitem [{\citenamefont {Shovkovy}(2013)}]{Shovkovy2013}%
  \BibitemOpen
  \bibfield  {author} {\bibinfo {author} {\bibfnamefont {I.~A.}\ \bibnamefont {Shovkovy}},\ }\bibinfo {title} {Magnetic catalysis: A review},\ in\ \href {https://doi.org/10.1007/978-3-642-37305-3_2} {\emph {\bibinfo {booktitle} {Strongly Interacting Matter in Magnetic Fields}}},\ \bibinfo {editor} {edited by\ \bibinfo {editor} {\bibfnamefont {D.}~\bibnamefont {Kharzeev}}, \bibinfo {editor} {\bibfnamefont {K.}~\bibnamefont {Landsteiner}}, \bibinfo {editor} {\bibfnamefont {A.}~\bibnamefont {Schmitt}},\ and\ \bibinfo {editor} {\bibfnamefont {H.-U.}\ \bibnamefont {Yee}}}\ (\bibinfo  {publisher} {Springer Berlin Heidelberg},\ \bibinfo {address} {Berlin, Heidelberg},\ \bibinfo {year} {2013})\ pp.\ \bibinfo {pages} {13--49}\BibitemShut {NoStop}%
\bibitem [{\citenamefont {Gusynin}\ \emph {et~al.}(1996)\citenamefont {Gusynin}, \citenamefont {Miransky},\ and\ \citenamefont {Shovkovy}}]{Gusynin:1995nb}%
  \BibitemOpen
  \bibfield  {author} {\bibinfo {author} {\bibfnamefont {V.}~\bibnamefont {Gusynin}}, \bibinfo {author} {\bibfnamefont {V.}~\bibnamefont {Miransky}},\ and\ \bibinfo {author} {\bibfnamefont {I.}~\bibnamefont {Shovkovy}},\ }\bibfield  {title} {\bibinfo {title} {{Dimensional reduction and catalysis of dynamical symmetry breaking by a magnetic field}},\ }\href {https://doi.org/10.1016/0550-3213(96)00021-1} {\bibfield  {journal} {\bibinfo  {journal} {Nucl. Phys. B}\ }\textbf {\bibinfo {volume} {462}},\ \bibinfo {pages} {249} (\bibinfo {year} {1996})},\ \Eprint {https://arxiv.org/abs/hep-ph/9509320} {arXiv:hep-ph/9509320} \BibitemShut {NoStop}%
\bibitem [{\citenamefont {Endr{\H{o}}di}\ \emph {et~al.}(2019)\citenamefont {Endr{\H{o}}di}, \citenamefont {Giordano}, \citenamefont {Katz}, \citenamefont {Kovacs},\ and\ \citenamefont {Pittler}}]{endrHodi2019magnetic}%
  \BibitemOpen
  \bibfield  {author} {\bibinfo {author} {\bibfnamefont {G.}~\bibnamefont {Endr{\H{o}}di}}, \bibinfo {author} {\bibfnamefont {M.}~\bibnamefont {Giordano}}, \bibinfo {author} {\bibfnamefont {S.~D.}\ \bibnamefont {Katz}}, \bibinfo {author} {\bibfnamefont {T.~G.}\ \bibnamefont {Kovacs}},\ and\ \bibinfo {author} {\bibfnamefont {F.}~\bibnamefont {Pittler}},\ }\bibfield  {title} {\bibinfo {title} {Magnetic catalysis and inverse catalysis for heavy pions},\ }\href {https://link.springer.com/article/10.1007/JHEP07(2019)007} {\bibfield  {journal} {\bibinfo  {journal} {Journal of high energy physics}\ }\textbf {\bibinfo {volume} {2019}},\ \bibinfo {pages} {1} (\bibinfo {year} {2019})}\BibitemShut {NoStop}%
\bibitem [{\citenamefont {Bocquet}\ \emph {et~al.}(1995)\citenamefont {Bocquet}, \citenamefont {Bonazzola}, \citenamefont {Gourgoulhon},\ and\ \citenamefont {Novak}}]{bocquet1995rotatingneutronstarmodels}%
  \BibitemOpen
  \bibfield  {author} {\bibinfo {author} {\bibfnamefont {M.}~\bibnamefont {Bocquet}}, \bibinfo {author} {\bibfnamefont {S.}~\bibnamefont {Bonazzola}}, \bibinfo {author} {\bibfnamefont {E.}~\bibnamefont {Gourgoulhon}},\ and\ \bibinfo {author} {\bibfnamefont {J.}~\bibnamefont {Novak}},\ }\href {https://arxiv.org/abs/gr-qc/9503044} {\bibinfo {title} {Rotating neutron star models with magnetic field}} (\bibinfo {year} {1995}),\ \Eprint {https://arxiv.org/abs/gr-qc/9503044} {arXiv:gr-qc/9503044 [gr-qc]} \BibitemShut {NoStop}%
\bibitem [{\citenamefont {Baym}\ \emph {et~al.}(1996)\citenamefont {Baym}, \citenamefont {B\"odeker},\ and\ \citenamefont {McLerran}}]{PhysRevD.53.662}%
  \BibitemOpen
  \bibfield  {author} {\bibinfo {author} {\bibfnamefont {G.}~\bibnamefont {Baym}}, \bibinfo {author} {\bibfnamefont {D.}~\bibnamefont {B\"odeker}},\ and\ \bibinfo {author} {\bibfnamefont {L.}~\bibnamefont {McLerran}},\ }\bibfield  {title} {\bibinfo {title} {Magnetic fields produced by phase transition bubbles in the electroweak phase transition},\ }\href {https://doi.org/10.1103/PhysRevD.53.662} {\bibfield  {journal} {\bibinfo  {journal} {Phys. Rev. D}\ }\textbf {\bibinfo {volume} {53}},\ \bibinfo {pages} {662} (\bibinfo {year} {1996})}\BibitemShut {NoStop}%
\bibitem [{\citenamefont {Tuchin}(2010)}]{PhysRevC.82.034904}%
  \BibitemOpen
  \bibfield  {author} {\bibinfo {author} {\bibfnamefont {K.}~\bibnamefont {Tuchin}},\ }\bibfield  {title} {\bibinfo {title} {Synchrotron radiation by fast fermions in heavy-ion collisions},\ }\href {https://doi.org/10.1103/PhysRevC.82.034904} {\bibfield  {journal} {\bibinfo  {journal} {Phys. Rev. C}\ }\textbf {\bibinfo {volume} {82}},\ \bibinfo {pages} {034904} (\bibinfo {year} {2010})}\BibitemShut {NoStop}%
\bibitem [{\citenamefont {Tuchin}(2013{\natexlab{b}})}]{Kiril2013ahep}%
  \BibitemOpen
  \bibfield  {author} {\bibinfo {author} {\bibfnamefont {K.}~\bibnamefont {Tuchin}},\ }\bibfield  {title} {\bibinfo {title} {Particle production in strong electromagnetic fields in relativistic heavy-ion collisions},\ }\href {https://doi.org/https://doi.org/10.1155/2013/490495} {\bibfield  {journal} {\bibinfo  {journal} {Advances in High Energy Physics}\ }\textbf {\bibinfo {volume} {2013}},\ \bibinfo {pages} {490495} (\bibinfo {year} {2013}{\natexlab{b}})}\BibitemShut {NoStop}%
\bibitem [{\citenamefont {Stewart}\ and\ \citenamefont {Tuchin}(2021)}]{STEWART2021122308}%
  \BibitemOpen
  \bibfield  {author} {\bibinfo {author} {\bibfnamefont {E.}~\bibnamefont {Stewart}}\ and\ \bibinfo {author} {\bibfnamefont {K.}~\bibnamefont {Tuchin}},\ }\bibfield  {title} {\bibinfo {title} {Continuous evolution of electromagnetic field in heavy-ion collisions},\ }\href {https://doi.org/https://doi.org/10.1016/j.nuclphysa.2021.122308} {\bibfield  {journal} {\bibinfo  {journal} {Nuclear Physics A}\ }\textbf {\bibinfo {volume} {1016}},\ \bibinfo {pages} {122308} (\bibinfo {year} {2021})}\BibitemShut {NoStop}%
\bibitem [{\citenamefont {Ghosh}\ and\ \citenamefont {Kurian}(2023)}]{Ghosh:2022vjp}%
  \BibitemOpen
  \bibfield  {author} {\bibinfo {author} {\bibfnamefont {R.}~\bibnamefont {Ghosh}}\ and\ \bibinfo {author} {\bibfnamefont {M.}~\bibnamefont {Kurian}},\ }\bibfield  {title} {\bibinfo {title} {{Magnetic-field-dependent electric-charge transport in hadronic medium at finite temperature}},\ }\href {https://doi.org/10.1103/PhysRevC.107.034903} {\bibfield  {journal} {\bibinfo  {journal} {Phys. Rev. C}\ }\textbf {\bibinfo {volume} {107}},\ \bibinfo {pages} {034903} (\bibinfo {year} {2023})},\ \Eprint {https://arxiv.org/abs/2211.06729} {arXiv:2211.06729 [hep-ph]} \BibitemShut {NoStop}%
\bibitem [{\citenamefont {Kalikotay}\ \emph {et~al.}(2020)\citenamefont {Kalikotay}, \citenamefont {Ghosh}, \citenamefont {Chaudhuri}, \citenamefont {Roy},\ and\ \citenamefont {Sarkar}}]{PhysRevD.102.076007}%
  \BibitemOpen
  \bibfield  {author} {\bibinfo {author} {\bibfnamefont {P.}~\bibnamefont {Kalikotay}}, \bibinfo {author} {\bibfnamefont {S.}~\bibnamefont {Ghosh}}, \bibinfo {author} {\bibfnamefont {N.}~\bibnamefont {Chaudhuri}}, \bibinfo {author} {\bibfnamefont {P.}~\bibnamefont {Roy}},\ and\ \bibinfo {author} {\bibfnamefont {S.}~\bibnamefont {Sarkar}},\ }\bibfield  {title} {\bibinfo {title} {Medium effects on the electrical and hall conductivities of a hot and magnetized pion gas},\ }\href {https://doi.org/10.1103/PhysRevD.102.076007} {\bibfield  {journal} {\bibinfo  {journal} {Phys. Rev. D}\ }\textbf {\bibinfo {volume} {102}},\ \bibinfo {pages} {076007} (\bibinfo {year} {2020})}\BibitemShut {NoStop}%
\bibitem [{\citenamefont {Das}\ \emph {et~al.}(2020)\citenamefont {Das}, \citenamefont {Mishra},\ and\ \citenamefont {Mohapatra}}]{PhysRevD.102.014030}%
  \BibitemOpen
  \bibfield  {author} {\bibinfo {author} {\bibfnamefont {A.}~\bibnamefont {Das}}, \bibinfo {author} {\bibfnamefont {H.}~\bibnamefont {Mishra}},\ and\ \bibinfo {author} {\bibfnamefont {R.~K.}\ \bibnamefont {Mohapatra}},\ }\bibfield  {title} {\bibinfo {title} {Magneto-seebeck coefficient and nernst coefficient of a hot and dense hadron gas},\ }\href {https://doi.org/10.1103/PhysRevD.102.014030} {\bibfield  {journal} {\bibinfo  {journal} {Phys. Rev. D}\ }\textbf {\bibinfo {volume} {102}},\ \bibinfo {pages} {014030} (\bibinfo {year} {2020})}\BibitemShut {NoStop}%
\bibitem [{\citenamefont {Das}\ \emph {et~al.}(2019{\natexlab{b}})\citenamefont {Das}, \citenamefont {Mishra},\ and\ \citenamefont {Mohapatra}}]{Das:2019wjg}%
  \BibitemOpen
  \bibfield  {author} {\bibinfo {author} {\bibfnamefont {A.}~\bibnamefont {Das}}, \bibinfo {author} {\bibfnamefont {H.}~\bibnamefont {Mishra}},\ and\ \bibinfo {author} {\bibfnamefont {R.~K.}\ \bibnamefont {Mohapatra}},\ }\bibfield  {title} {\bibinfo {title} {{Electrical conductivity and Hall conductivity of a hot and dense hadron gas in a magnetic field: A relaxation time approach}},\ }\href {https://doi.org/10.1103/PhysRevD.99.094031} {\bibfield  {journal} {\bibinfo  {journal} {Phys. Rev. D}\ }\textbf {\bibinfo {volume} {99}},\ \bibinfo {pages} {094031} (\bibinfo {year} {2019}{\natexlab{b}})},\ \Eprint {https://arxiv.org/abs/1903.03938} {arXiv:1903.03938 [hep-ph]} \BibitemShut {NoStop}%
\bibitem [{\citenamefont {Dey}\ \emph {et~al.}(2025)\citenamefont {Dey}, \citenamefont {K.}, \citenamefont {Dash},\ and\ \citenamefont {Nandi}}]{Dey:2025hgw}%
  \BibitemOpen
  \bibfield  {author} {\bibinfo {author} {\bibfnamefont {D.}~\bibnamefont {Dey}}, \bibinfo {author} {\bibfnamefont {G.~K.}\ \bibnamefont {K.}}, \bibinfo {author} {\bibfnamefont {S.}~\bibnamefont {Dash}},\ and\ \bibinfo {author} {\bibfnamefont {B.~K.}\ \bibnamefont {Nandi}},\ }\bibfield  {title} {\bibinfo {title} {{Viscous effects of a hot QGP medium in a time dependent magnetic field and their phenomenological significance}},\ }\href {https://doi.org/10.1103/PhysRevD.111.096014} {\bibfield  {journal} {\bibinfo  {journal} {Phys. Rev. D}\ }\textbf {\bibinfo {volume} {111}},\ \bibinfo {pages} {096014} (\bibinfo {year} {2025})},\ \Eprint {https://arxiv.org/abs/2501.19349} {arXiv:2501.19349 [hep-ph]} \BibitemShut {NoStop}%
\bibitem [{\citenamefont {K}\ \emph {et~al.}(2021)\citenamefont {K}, \citenamefont {Kurian},\ and\ \citenamefont {Chandra}}]{K:2021sct}%
  \BibitemOpen
  \bibfield  {author} {\bibinfo {author} {\bibfnamefont {G.~K.}\ \bibnamefont {K}}, \bibinfo {author} {\bibfnamefont {M.}~\bibnamefont {Kurian}},\ and\ \bibinfo {author} {\bibfnamefont {V.}~\bibnamefont {Chandra}},\ }\bibfield  {title} {\bibinfo {title} {{Electromagnetic response of hot QCD medium in the presence of background time-varying fields}},\ }\href {https://doi.org/10.1103/PhysRevD.104.094037} {\bibfield  {journal} {\bibinfo  {journal} {Phys. Rev. D}\ }\textbf {\bibinfo {volume} {104}},\ \bibinfo {pages} {094037} (\bibinfo {year} {2021})},\ \Eprint {https://arxiv.org/abs/2108.06791} {arXiv:2108.06791 [hep-ph]} \BibitemShut {NoStop}%
\bibitem [{\citenamefont {Satow}(2014)}]{Satow:2014lia}%
  \BibitemOpen
  \bibfield  {author} {\bibinfo {author} {\bibfnamefont {D.}~\bibnamefont {Satow}},\ }\bibfield  {title} {\bibinfo {title} {{Nonlinear electromagnetic response in quark-gluon plasma}},\ }\href {https://doi.org/10.1103/PhysRevD.90.034018} {\bibfield  {journal} {\bibinfo  {journal} {Phys. Rev. D}\ }\textbf {\bibinfo {volume} {90}},\ \bibinfo {pages} {034018} (\bibinfo {year} {2014})},\ \Eprint {https://arxiv.org/abs/1406.7032} {arXiv:1406.7032 [hep-ph]} \BibitemShut {NoStop}%
\bibitem [{\citenamefont {Mitra}\ and\ \citenamefont {Chandra}(2018)}]{PhysRevD.97.034032}%
  \BibitemOpen
  \bibfield  {author} {\bibinfo {author} {\bibfnamefont {S.}~\bibnamefont {Mitra}}\ and\ \bibinfo {author} {\bibfnamefont {V.}~\bibnamefont {Chandra}},\ }\bibfield  {title} {\bibinfo {title} {Covariant kinetic theory for effective fugacity quasiparticle model and first order transport coefficients for hot qcd matter},\ }\href {https://doi.org/10.1103/PhysRevD.97.034032} {\bibfield  {journal} {\bibinfo  {journal} {Phys. Rev. D}\ }\textbf {\bibinfo {volume} {97}},\ \bibinfo {pages} {034032} (\bibinfo {year} {2018})}\BibitemShut {NoStop}%
\bibitem [{\citenamefont {Groot}\ \emph {et~al.}(1980)\citenamefont {Groot}, \citenamefont {Leeuwen}, \citenamefont {van Weert},\ and\ \citenamefont {Weert}}]{groot1980relativistic}%
  \BibitemOpen
  \bibfield  {author} {\bibinfo {author} {\bibfnamefont {S.~R.}\ \bibnamefont {Groot}}, \bibinfo {author} {\bibfnamefont {W.~A.}\ \bibnamefont {Leeuwen}}, \bibinfo {author} {\bibfnamefont {C.~G.}\ \bibnamefont {van Weert}},\ and\ \bibinfo {author} {\bibfnamefont {C.~G.}\ \bibnamefont {Weert}},\ }\href@noop {} {\emph {\bibinfo {title} {Relativistic kinetic theory: principles and applications}}}\ (\bibinfo  {publisher} {North Holland},\ \bibinfo {year} {1980})\BibitemShut {NoStop}%
\bibitem [{\citenamefont {Gowthama}\ \emph {et~al.}(2022)\citenamefont {Gowthama}, \citenamefont {Kurian},\ and\ \citenamefont {Chandra}}]{PhysRevD.106.034008}%
  \BibitemOpen
  \bibfield  {author} {\bibinfo {author} {\bibfnamefont {K.~K.}\ \bibnamefont {Gowthama}}, \bibinfo {author} {\bibfnamefont {M.}~\bibnamefont {Kurian}},\ and\ \bibinfo {author} {\bibfnamefont {V.}~\bibnamefont {Chandra}},\ }\bibfield  {title} {\bibinfo {title} {Thermal and thermoelectric responses of hot qcd medium in time-varying magnetic fields},\ }\href {https://doi.org/10.1103/PhysRevD.106.034008} {\bibfield  {journal} {\bibinfo  {journal} {Phys. Rev. D}\ }\textbf {\bibinfo {volume} {106}},\ \bibinfo {pages} {034008} (\bibinfo {year} {2022})}\BibitemShut {NoStop}%
\bibitem [{\citenamefont {Mitra}\ and\ \citenamefont {Sarkar}(2014)}]{PhysRevD.89.054013}%
  \BibitemOpen
  \bibfield  {author} {\bibinfo {author} {\bibfnamefont {S.}~\bibnamefont {Mitra}}\ and\ \bibinfo {author} {\bibfnamefont {S.}~\bibnamefont {Sarkar}},\ }\bibfield  {title} {\bibinfo {title} {Medium effects on the thermal conductivity of a hot pion gas},\ }\href {https://doi.org/10.1103/PhysRevD.89.054013} {\bibfield  {journal} {\bibinfo  {journal} {Phys. Rev. D}\ }\textbf {\bibinfo {volume} {89}},\ \bibinfo {pages} {054013} (\bibinfo {year} {2014})}\BibitemShut {NoStop}%
\bibitem [{\citenamefont {Mitra}\ and\ \citenamefont {Chandra}(2017)}]{PhysRevD.96.094003}%
  \BibitemOpen
  \bibfield  {author} {\bibinfo {author} {\bibfnamefont {S.}~\bibnamefont {Mitra}}\ and\ \bibinfo {author} {\bibfnamefont {V.}~\bibnamefont {Chandra}},\ }\bibfield  {title} {\bibinfo {title} {Transport coefficients of a hot qcd medium and their relative significance in heavy-ion collisions},\ }\href {https://doi.org/10.1103/PhysRevD.96.094003} {\bibfield  {journal} {\bibinfo  {journal} {Phys. Rev. D}\ }\textbf {\bibinfo {volume} {96}},\ \bibinfo {pages} {094003} (\bibinfo {year} {2017})}\BibitemShut {NoStop}%
\bibitem [{\citenamefont {Bhatt}\ \emph {et~al.}(2019)\citenamefont {Bhatt}, \citenamefont {Das},\ and\ \citenamefont {Mishra}}]{Bhatt:2018ncr}%
  \BibitemOpen
  \bibfield  {author} {\bibinfo {author} {\bibfnamefont {J.~R.}\ \bibnamefont {Bhatt}}, \bibinfo {author} {\bibfnamefont {A.}~\bibnamefont {Das}},\ and\ \bibinfo {author} {\bibfnamefont {H.}~\bibnamefont {Mishra}},\ }\bibfield  {title} {\bibinfo {title} {{Thermoelectric effect and Seebeck coefficient for hot and dense hadronic matter}},\ }\href {https://doi.org/10.1103/PhysRevD.99.014015} {\bibfield  {journal} {\bibinfo  {journal} {Phys. Rev. D}\ }\textbf {\bibinfo {volume} {99}},\ \bibinfo {pages} {014015} (\bibinfo {year} {2019})},\ \Eprint {https://arxiv.org/abs/1808.02789} {arXiv:1808.02789 [hep-ph]} \BibitemShut {NoStop}%
\bibitem [{\citenamefont {Rath}\ and\ \citenamefont {Dash}(2023)}]{rath2023effects}%
  \BibitemOpen
  \bibfield  {author} {\bibinfo {author} {\bibfnamefont {S.}~\bibnamefont {Rath}}\ and\ \bibinfo {author} {\bibfnamefont {S.}~\bibnamefont {Dash}},\ }\bibfield  {title} {\bibinfo {title} {Effects of weak magnetic field and finite chemical potential on the transport of charge and heat in hot qcd matter},\ }\href {https://link.springer.com/article/10.1140/epja/s10050-023-00941-9} {\bibfield  {journal} {\bibinfo  {journal} {The European Physical Journal A}\ }\textbf {\bibinfo {volume} {59}},\ \bibinfo {pages} {25} (\bibinfo {year} {2023})}\BibitemShut {NoStop}%
\bibitem [{\citenamefont {Bhalerao}\ \emph {et~al.}(2005)\citenamefont {Bhalerao}, \citenamefont {Blaizot}, \citenamefont {Borghini},\ and\ \citenamefont {Ollitrault}}]{Bhalerao:2005mm}%
  \BibitemOpen
  \bibfield  {author} {\bibinfo {author} {\bibfnamefont {R.~S.}\ \bibnamefont {Bhalerao}}, \bibinfo {author} {\bibfnamefont {J.-P.}\ \bibnamefont {Blaizot}}, \bibinfo {author} {\bibfnamefont {N.}~\bibnamefont {Borghini}},\ and\ \bibinfo {author} {\bibfnamefont {J.-Y.}\ \bibnamefont {Ollitrault}},\ }\bibfield  {title} {\bibinfo {title} {{Elliptic flow and incomplete equilibration at RHIC}},\ }\href {https://doi.org/10.1016/j.physletb.2005.08.131} {\bibfield  {journal} {\bibinfo  {journal} {Phys. Lett. B}\ }\textbf {\bibinfo {volume} {627}},\ \bibinfo {pages} {49} (\bibinfo {year} {2005})},\ \Eprint {https://arxiv.org/abs/nucl-th/0508009} {arXiv:nucl-th/0508009} \BibitemShut {NoStop}%
\bibitem [{\citenamefont {Rath}\ and\ \citenamefont {Dash}(2022)}]{rath2022momentum}%
  \BibitemOpen
  \bibfield  {author} {\bibinfo {author} {\bibfnamefont {S.}~\bibnamefont {Rath}}\ and\ \bibinfo {author} {\bibfnamefont {S.}~\bibnamefont {Dash}},\ }\bibfield  {title} {\bibinfo {title} {Momentum transport properties of a hot and dense qcd matter in a weak magnetic field},\ }\href {https://link.springer.com/article/10.1140/epjc/s10052-022-10757-4} {\bibfield  {journal} {\bibinfo  {journal} {The European Physical Journal C}\ }\textbf {\bibinfo {volume} {82}},\ \bibinfo {pages} {797} (\bibinfo {year} {2022})}\BibitemShut {NoStop}%
\bibitem [{\citenamefont {Chakraborty}\ and\ \citenamefont {Kapusta}(2011)}]{PhysRevC.83.014906}%
  \BibitemOpen
  \bibfield  {author} {\bibinfo {author} {\bibfnamefont {P.}~\bibnamefont {Chakraborty}}\ and\ \bibinfo {author} {\bibfnamefont {J.~I.}\ \bibnamefont {Kapusta}},\ }\bibfield  {title} {\bibinfo {title} {Quasiparticle theory of shear and bulk viscosities of hadronic matter},\ }\href {https://doi.org/10.1103/PhysRevC.83.014906} {\bibfield  {journal} {\bibinfo  {journal} {Phys. Rev. C}\ }\textbf {\bibinfo {volume} {83}},\ \bibinfo {pages} {014906} (\bibinfo {year} {2011})}\BibitemShut {NoStop}%
\bibitem [{\citenamefont {{PHENIX Collaboration}}(2003)}]{PhysRevLett.91.182301}%
  \BibitemOpen
  \bibfield  {author} {\bibinfo {author} {\bibnamefont {{PHENIX Collaboration}}},\ }\bibfield  {title} {\bibinfo {title} {Elliptic flow of identified hadrons in $\mathrm{A}\mathrm{u}+\mathrm{A}\mathrm{u}$ collisions at $\sqrt{{s}_{NN}}=200\text{ }\text{ }\mathrm{G}\mathrm{e}\mathrm{V}$},\ }\href {https://doi.org/10.1103/PhysRevLett.91.182301} {\bibfield  {journal} {\bibinfo  {journal} {Phys. Rev. Lett.}\ }\textbf {\bibinfo {volume} {91}},\ \bibinfo {pages} {182301} (\bibinfo {year} {2003})}\BibitemShut {NoStop}%
\bibitem [{\citenamefont {Gombeaud}\ and\ \citenamefont {Ollitrault}(2008)}]{Gombeaud:2007ub}%
  \BibitemOpen
  \bibfield  {author} {\bibinfo {author} {\bibfnamefont {C.}~\bibnamefont {Gombeaud}}\ and\ \bibinfo {author} {\bibfnamefont {J.-Y.}\ \bibnamefont {Ollitrault}},\ }\bibfield  {title} {\bibinfo {title} {{Elliptic flow in transport theory and hydrodynamics}},\ }\href {https://doi.org/10.1103/PhysRevC.77.054904} {\bibfield  {journal} {\bibinfo  {journal} {Phys. Rev. C}\ }\textbf {\bibinfo {volume} {77}},\ \bibinfo {pages} {054904} (\bibinfo {year} {2008})},\ \Eprint {https://arxiv.org/abs/nucl-th/0702075} {arXiv:nucl-th/0702075} \BibitemShut {NoStop}%
\bibitem [{\citenamefont {Kalikotay}\ \emph {et~al.}(2024)\citenamefont {Kalikotay}, \citenamefont {Ghosh}, \citenamefont {Chaudhuri}, \citenamefont {Roy},\ and\ \citenamefont {Sarkar}}]{kalikotay2024electrical}%
  \BibitemOpen
  \bibfield  {author} {\bibinfo {author} {\bibfnamefont {P.}~\bibnamefont {Kalikotay}}, \bibinfo {author} {\bibfnamefont {S.}~\bibnamefont {Ghosh}}, \bibinfo {author} {\bibfnamefont {N.}~\bibnamefont {Chaudhuri}}, \bibinfo {author} {\bibfnamefont {P.}~\bibnamefont {Roy}},\ and\ \bibinfo {author} {\bibfnamefont {S.}~\bibnamefont {Sarkar}},\ }\bibfield  {title} {\bibinfo {title} {Electrical conductivity and shear viscosity of a pion gas in a thermo-magnetic medium},\ }\href {https://link.springer.com/article/10.1140/epja/s10050-024-01291-w} {\bibfield  {journal} {\bibinfo  {journal} {The European Physical Journal A}\ }\textbf {\bibinfo {volume} {60}},\ \bibinfo {pages} {71} (\bibinfo {year} {2024})}\BibitemShut {NoStop}%
\bibitem [{\citenamefont {Kumar~Singh}\ \emph {et~al.}(2025)\citenamefont {Kumar~Singh}, \citenamefont {Bhadury}, \citenamefont {Kurian},\ and\ \citenamefont {Chandra}}]{KumarSingh:2025kml}%
  \BibitemOpen
  \bibfield  {author} {\bibinfo {author} {\bibfnamefont {S.}~\bibnamefont {Kumar~Singh}}, \bibinfo {author} {\bibfnamefont {S.}~\bibnamefont {Bhadury}}, \bibinfo {author} {\bibfnamefont {M.}~\bibnamefont {Kurian}},\ and\ \bibinfo {author} {\bibfnamefont {V.}~\bibnamefont {Chandra}},\ }\bibfield  {title} {\bibinfo {title} {{Particle number diffusion in second-order relativistic dissipative hydrodynamics with momentum-dependent relaxation time}},\ }\href {https://doi.org/10.1103/sf5m-87w7} {\bibfield  {journal} {\bibinfo  {journal} {Phys. Rev. D}\ }\textbf {\bibinfo {volume} {111}},\ \bibinfo {pages} {114007} (\bibinfo {year} {2025})},\ \Eprint {https://arxiv.org/abs/2501.17442} {arXiv:2501.17442 [hep-ph]} \BibitemShut {NoStop}%
\bibitem [{\citenamefont {Singh}\ \emph {et~al.}(2024)\citenamefont {Singh}, \citenamefont {Kurian},\ and\ \citenamefont {Chandra}}]{Singh:2024leo}%
  \BibitemOpen
  \bibfield  {author} {\bibinfo {author} {\bibfnamefont {S.~K.}\ \bibnamefont {Singh}}, \bibinfo {author} {\bibfnamefont {M.}~\bibnamefont {Kurian}},\ and\ \bibinfo {author} {\bibfnamefont {V.}~\bibnamefont {Chandra}},\ }\bibfield  {title} {\bibinfo {title} {{Revisiting shear stress tensor evolution: Nonresistive magnetohydrodynamics with momentum-dependent relaxation time}},\ }\href {https://doi.org/10.1103/PhysRevD.110.014004} {\bibfield  {journal} {\bibinfo  {journal} {Phys. Rev. D}\ }\textbf {\bibinfo {volume} {110}},\ \bibinfo {pages} {014004} (\bibinfo {year} {2024})},\ \Eprint {https://arxiv.org/abs/2403.13160} {arXiv:2403.13160 [hep-ph]} \BibitemShut {NoStop}%
\end{thebibliography}%

\end{document}